\documentclass[11pt]{article}
\usepackage{jheppub}
\usepackage{amsmath,amssymb,amsfonts,amsthm,amscd,mathrsfs,dsfont}
\usepackage{xcolor}
\usepackage{tikz}
\usetikzlibrary{shapes,shapes.misc,snakes,decorations.pathreplacing,decorations.markings}

\newcommand{\be}{\begin{equation}}
\newcommand{\ee}{\end{equation}}

\newcommand{\bea}{\begin{eqnarray}}
\newcommand{\eea}{\end{eqnarray}}

\newcommand{\ba}{\begin{equation}\begin{aligned}}
\newcommand{\ea}{\end{aligned}\end{equation}}

\newcommand{\Df}{\Delta_{\phi}}
\newcommand{\Dp}{\Delta_{\psi}}

\newcommand{\cA}{\mathcal{A}}
\newcommand{\cC}{\mathcal{C}}
\newcommand{\cD}{\mathcal{D}}
\newcommand{\cO}{\mathcal{O}}
\newcommand{\cM}{\mathcal{M}}

\newcommand{\cG}{\mathcal{G}}

\newcommand{\dDisc}{\mathrm{dDisc}}
\newcommand{\Dgap}{\Delta_{\textrm{gap}}}

\newcommand{\mS}{\mathrm{s}}
\newcommand{\mT}{\mathrm{t}}
\newcommand{\mU}{\mathrm{u}}

\newcommand{\rB}{\mathrm{B}}
\newcommand{\rF}{\mathrm{F}}

\usepackage{todonotes}

\numberwithin{equation}{section}

\author[1]{Waltraut Knop,}
\author[2]{Dalimil Maz\'{a}\v{c}}
\affiliation[1]{C. N. Yang Institute for Theoretical Physics, SUNY, Stony Brook, NY 11794, USA}
\affiliation[2]{School of Natural Sciences, Institute for Advanced Study, Princeton, NJ 08540, USA}
\emailAdd{waltraut.knop@stonybrook.edu}
\emailAdd{dmazac@ias.edu}

\preprint{YITP-SB-2022-13}

\begin{document}

\title{Dispersive Sum Rules in AdS$_{\text{2}}$}

\abstract{Dispersion relations for S-matrices and CFT correlators translate UV consistency into bounds on IR observables. In this note, we construct dispersive sum rules for 1D CFTs. We use them to prove bounds on higher-derivative couplings in weakly-coupled non-gravitational EFTs in AdS$_2$. At the leading order in the bulk-point limit, the bounds agree with the flat-space result. We compute the leading universal effect of finite AdS radius on the bounds. Along the way, we give an explicit formula for anomalous dimensions in general higher-derivative contact Witten diagrams in AdS$_2$.
}

\maketitle

\section{Introduction}
One of the most powerful ideas in theoretical physics is the systematic application of universal principles, such as symmetry, causality and unitarity, to constrain observable quantities. This approach, generally referred to as the bootstrap, has led to many insights into the dynamics of quantum field theory and quantum gravity, see \cite{Poland:2018epd,Hartman:2022zik,Poland:2022qrs,Gopakumar:2022kof,Kruczenski:2022lot} and references therein.

The precise form of observables and thus also the implementation of the bootstrap strategy depends on the global geometry of spacetime. For example, in asymptotically flat space, a relevant set of observables is the S-matrix. In asymptotically AdS space, it is the correlators of the boundary CFT. In this work, we will focus on scattering in 1+1 dimensions. We will use the conformal bootstrap in 1D CFTs to constrain scattering in QFTs in AdS$_2$. We will also draw parallels with the analogous results about 2D S-matrices in flat space.

In principle, the ultimate goal of the S-matrix bootstrap is to construct the most general Lorentz-invariant S-matrix consistent with unitarity and causality. Needless to say, the solution of this problem is currently far out of reach. On the other hand, if we only demand consistency order by order at small energy, then the complete solution is known. The solution is called low-energy effective field theory (EFT). Indeed, the purpose of low-energy EFT is to provide a parametrization of the most general S-matrix consistent at low energy \cite{Weinberg:1978kz,Weinberg:1995mt,Weinberg:1996kr}. The situation is similar for scattering in AdS. We can study solutions of the conformal bootstrap equations in the boundary CFT order by order in an expansion around mean field theory, i.e.\ free theory in AdS. Again, the most general perturbative solution of the bootstrap is parametrized by a bulk effective Lagrangian and computed using bulk Witten diagrams \cite{Heemskerk:2009pn,Fitzpatrick:2010zm,Fitzpatrick:2011dm}.

However, there is no guarantee that a perturbative solution of the bootstrap equations arises from a theory consistent at all energies. In other words, the set of solutions of the full bootstrap could be much smaller than the set of EFTs. Indeed, it has long been known that low-energy couplings must satisfy various inequalities if the EFT arises from a UV-complete theory \cite{Pham:1985cr,Adams:2006sv}. Such bounds have recently received a renewed attention, see for example \cite{Camanho:2014apa,Hartman:2015lfa,Baumann:2015nta,Bellazzini:2015cra,Bonifacio:2016wcb,deRham:2017avq,Caron-Huot:2022ugt}. A central task facing the modern S-matrix and conformal bootstrap is to find the complete set of constraints on the low-energy parameters arising in fully consistent theories.

A fruitful approach to deriving constraints on IR observables following from UV consistency stems from dispersion relations. Dispersion relations distill causality and unitarity into precise sum rules satisfied by the dynamical data. In flat space, dispersion relations follow from analyticity and boundedness of the S-matrix at high energy by a contour deformation in the space of complexified momenta. 

Recently, dispersive sum rules for the S-matrix were explored systematically in \cite{Bellazzini:2020cot,Arkani-Hamed:2020blm,Tolley:2020gtv,Caron-Huot:2020cmc,Caron-Huot:2021rmr}. Following \cite{Bellazzini:2020cot}, one can introduce the arcs, which are certain integrals of the two-to-two S-matrix defined at a variable energy scale $M$. On one hand, the arcs are computable in terms of a low-energy EFT valid at the scale $M$. On the other hand, a contour deformation expresses the arcs as moments of the imaginary part of the S-matrix at energies above $M$. UV unitarity leads to positivity properties of these moments, yielding inequalities satisfied by the IR parameters. An interesting outcome of these inequalities is a justification of the expectation that higher-derivatives couplings in the EFT should be suppressed by inverse powers of the UV cutoff $M$ dictated by dimensional analysis.

Dispersive arguments have a natural implementation for scattering in AdS in the language of the conformal bootstrap. Indeed, causality and unitarity are deeply embedded in the conformal bootstrap. In particular, the CFT crossing equation is a statement of causality, expressing the commutativity of spacelike-separated local operators in a four-point function. Dispersive CFT sum rules can be derived as rigorous consequences of the CFT crossing equation in Lorentzian signature by applying suitable linear functionals to it \cite{Carmi:2019cub,Mazac:2019shk}. The defining property of a dispersive functional is that heavy operators enter only through the double commutator aka double discontinuity. This statement is the AdS analogue of the idea that heavy states enter the dispersive sum rules in flat space only through the imaginary part of the S-matrix.

Dispersive CFT sum rules have appeared in various forms in the conformal bootstrap literature, for example as superconvergence sum rules \cite{Kologlu:2019bco}, analytic extremal functionals \cite{Mazac:2016qev,Mazac:2018mdx,Mazac:2018ycv,Paulos:2019gtx}, or as sum rules arising in the Polyakov-Mellin approach to the conformal bootstrap \cite{Gopakumar:2016cpb,Penedones:2019tng,Carmi:2020ekr,Gopakumar:2021dvg}. In the context of CFTs in $d\geq 2$ dimensions, they were studied systematically in \cite{Caron-Huot:2020adz}.

In analogy with flat space, these sum rules can be used to prove bounds on couplings in a low-energy EFT in AdS. The bounds follow by splitting the contributions to the CFT crossing equation into those of light and heavy operators, where the separation occurs at a scaling dimension $\Dgap\sim M R_{\text{AdS}}$. The contribution of the light operators to a dispersive sum rule defines the AdS analogue of the arc variables. These arcs can be computed in terms of an EFT in AdS assuming this EFT is valid up to the scale $\Dgap$. The crossing equation relates the arcs to moments of the contribution of heavy operators to the double commutator. Positivity of these contributions then implies bounds on the arcs and hence on the IR observables. This logic was used in \cite{Caron-Huot:2021enk} to prove bounds on higher-derivative couplings in weakly-coupled gravitational theories in AdS$_D$ with $D\geq 4$, building on the flat-space arguments of \cite{Caron-Huot:2021rmr}. It was shown that at the leading order at large $\Dgap$, the bounds are at least as strong as the corresponding bounds in flat space. In particular, there remains the logical possibility that the AdS bounds are strictly stronger than the flat space bounds.

The main purpose of this note is to extend dispersive arguments to AdS$_2$ and deduce EFT bounds in that case. We imagine placing a non-gravitational QFT in AdS$_2$ and study it through the lens of the boundary correlators, as advocated in \cite{Paulos:2016fap} and further explored more recently in \cite{Antunes:2021abs}. The latter reference also derives a bound on the leading irrelevant interaction in a scalar theory.

The set of boundary correlators defines what one may call a 1D CFT. While not including a stress tensor, such 1D CFT satisfies a natural extension of the higher-$d$ conformal bootstrap equations to $d=1$. The 1D conformal bootstrap is the simplest nontrivial case of the conformal bootstrap. It allows one to study the conformal bootstrap equations in the simplest possible setting and has already lead to many insights \cite{Liendo:2018ukf,Mazac:2018qmi,Hartman:2019pcd,Paulos:2020zxx,Ferrero:2021bsb}. It is therefore natural to look for the analogue of dispersive bounds in this case. In particular, we will address the question of optimality of the AdS bounds by showing that at the leading order at large $\Dgap$, they exactly agree with the flat-space bounds.

The main difference between $d>1$ and $d=1$ is the absence of a transverse impact parameter space in the bulk in the latter case. This considerably simplifies kinematics, but does not limit the power of the bootstrap in constraining bulk scattering. 

Our central technical result is an explicit construction of dispersive sum rules $C_k$ for 1D CFTs. These play the same role for AdS$_2$ scattering as the spin-$k$ subtracted dispersion relation plays for the 2D S-matrix. The construction of $C_k$ follows the template of 1D analytic extremal functionals of \cite{Mazac:2016qev,Mazac:2018mdx,Mazac:2018ycv}. $C_k$ is defined as a weighted integral of the 1D conformal bootstrap equation. A contour deformation then shows that heavy operators enter $C_k$ only though their double discontinuity. The parameter $k$ controls the inverse power of $\Delta$ suppressing the heavy contributions.

We will define the AdS$_2$ arc variables as the contribution of operators with $\Delta<\Dgap$ to the dispersive sum rule $C_k$. This definition makes sense without resorting to any EFT description of bulk physics. Assuming bulk physics is well-approximated by a tree-level EFT valid up to scale $\Dgap$, the arc variables defined by $C_k$ compute precisely the quartic higher-derivative couplings.

The logic of dispersive sum rules explained above then leads to bounds on the higher-derivative couplings. The bounds are nontrivial functions of $\Dgap$. This is because by placing a theory in AdS$_2$ with finite radius, we introduce a new dimensionless parameter into the problem. $\Dgap$ is this parameter. We will see that in the regime $\Dgap\gg1$, our bounds agree with the flat space results. In particular, the couplings are suppressed by the expected inverse powers of $\Dgap$. Unlike in the case of higher dimensions, it will be manifest that the AdS$_{2}$ bound can not be improved beyond the flat-space bound (at the leading order in $1/\Dgap$) without considering other correlation functions. We will also compute the universal leading correction to the bounds coming from finite size of AdS. 

The rest of this note is organized as follows. In Section \ref{sec:SMatrix}, we review dispersive arguments in the context of scattering in 2D flat space. In Section \ref{sec:CFT}, we review the kinematics of AdS$_2$/CFT$_1$, give a general construction of dispersive sum rules and construct higher-derivative contact Witten diagrams. Our main results appear in Section \ref{sec:Dispersion}, where we construct the $C_k$ sum rules, and use them to prove bounds on higher-derivative couplings in AdS$_2$. We conclude in Section \ref{sec:conclusions}. Appendix \ref{app:gammas} includes an explicit formula for the anomalous dimensions in higher-derivative contact Witten diagrams in AdS$_2$.

\vspace{15pt}
\noindent\textbf{Note added:} We are coordinating our submission with \cite{Cordova:2022pbl}, which studies complementary aspects of scattering in 1+1 dimensions using the conformal bootstrap in 1D CFTs.

\section{Flat space bounds from dispersion relations}\label{sec:SMatrix}
\subsection{2D S-matrices}
We will start by reviewing several well-known facts about relativistic two-to-two scattering in 1+1 dimensions. Particles in 1+1 dimensions do not carry spin, but can possess either bosonic or fermionic statistics. We will consider both possibilities.

Let $|p\rangle$ be the one-particle state of a stable particle of mass $m$ and spatial momentum $p$. The states carry the standard relativistic normalization
\be
\langle q | p \rangle = 4\pi E\,\delta(p-q)\,,
\ee
where $E(p) = \sqrt{p^2+m^2}$. One can define the scattering in states and out states of two identical particles $|p_1,p_2\rangle_{\text{in}}$, $|p_1,p_2\rangle_{\text{out}}$. They satisfy $|p_1,p_2\rangle_{\text{in}} = \pm |p_2,p_1\rangle_{\text{in}}$, where here and in what follows, the upper and lower sign corresponds respectively to bosons and fermions. They are normalized as tensor products of one-particle states
\be
{}_{\text{in}}\!\langle p_3,p_4 | p_1,p_2 \rangle_{\text{in}} = (4\pi)^2 E_1 E_2 [\delta(p_1-p_3)\delta(p_2-p_4) \pm \delta(p_1-p_4)\delta(p_2-p_3)]\,.
\ee

The S-matrix $\widehat{S}$ is the operator carrying out states to in states
\be
 |p_1,p_2\rangle_{\text{in}} = \widehat{S} |p_1,p_2\rangle_{\text{out}}\,.
\ee
It is standard to write $\widehat{S}$ as follows
\be
\widehat{S} = \mathds{1} + i \widehat{T}\,,
\ee
where $i \widehat{T}$ is the connected amplitude. We are interested in the matrix elements of $\widehat{S}$ between in states. Poincar\'{e} invariance implies that
\be
{}_{\text{in}}\!\langle p_3,p_4 |i \widehat{T}| p_1,p_2 \rangle_{\text{in}} = i(2\pi)^2\delta(E_1+E_2-E_3-E_4)\delta(p_1+p_2-p_3-p_4) T(p_i)\,,
\label{eq:connectedAmplitude}
\ee
where $T(p_i)$ is defined on the support of the delta functions, is Lorentz-invariant, and satisfies
\be
T(p_1,p_2,p_3,p_4) = \pm T(p_2,p_1,p_3,p_4) = \pm T(p_1,p_2,p_4,p_3) = T(p_2,p_1,p_4,p_3)\,.
\ee
The energy-momentum conserving delta functions in \eqref{eq:connectedAmplitude} are supported in the region ($p_1=p_3$ and $p_2=p_4$) or ($p_1=p_4$ and $p_2=p_3$). Thus two-to-two scattering is fully encoded in $T(p_1,p_2,p_1,p_2)$. Symmetry under $p_1\leftrightarrow p_2$ and Lorentz invariance imply
\be
T(p_1,p_2,p_1,p_2) = \cM(s)\,,
\ee
where
\be
s = -(p_1+p_2)^2\,,\qquad
t = -(p_1-p_4)^2 = 4m^2 - s\,,\qquad u = -(p_1-p_3)^2 = 0\,.
\ee
It follows that the S-matrix elements take the form
\be
{}_{\text{in}}\!\langle p_3,p_4 |\widehat{S}| p_1,p_2 \rangle_{\text{in}} = (4\pi)^2E_1E_2[\delta(p_1-p_3)\delta(p_2-p_4) \pm \delta(p_1-p_4)\delta(p_2-p_3)] S(s)\,,
\ee
where
\be
S(s) = 1 + \frac{i \cM(s)}{2\sqrt{s(s-4m^2)}}\,.\label{eq:SfromM}
\ee
The square root in the denominator comes from the Jacobian transforming between the two types of delta-functions.

Physical scattering occurs for $s \geq 4m^2$ but $S(s)$ admits an analytic continuation to the complex $s$-plane. The analytic continuation exhibits a pair of branch cuts for $s\geq 4m^2$ and $s\leq 0$, corresponding to s- and t-channel scattering. The physical scattering amplitude \eqref{eq:SfromM} is the limiting value of $S(s)$ as we approach the s-channel branch cut from $\mathrm{Im}(s)>0$. $S(s)$ is meromorphic away from these cuts. Its poles must all be simple and are only allowed to occur for real $s$ with $0<s<4m^2$. They correspond to bound states of the two scattered particles.

Crossing symmetry is the statement that $S(s)$ below the t-channel cut computes the physical scattering amplitude of particle-antiparticle scattering. In the rest of this note, we will assume that the scattered particles are their own antiparticles. In that case, crossing symmetry becomes the statement
\be
S(s) = S(4m^2 -s)\,.\label{eq:Ssym}
\ee
Note that there is no overall minus sign here even for identical fermions.

The S-matrix is a unitary operator
\be
\widehat{S}^{\dagger}\widehat{S} = \mathds{1}.
\ee
By computing the matrix element of this identity between two-particle in states, and inserting a complete set of in states between $\widehat{S}^{\dagger}$ and $\widehat{S}$, we recover the upper bound
\be
|S(s)|^2 \leq 1
\ee
valid in the physical region $s\in [4m^2,\infty)$. Using \eqref{eq:SfromM}, it becomes the following statement about $\cM(s)$
\be
\mathrm{Im}[\cM(s)] \geq   \frac{|\cM(s)|^2}{4\sqrt{s(s-4m^2)}}\,. \label{eq:UnitOnM}
\ee
In particular, $\mathrm{Im}[\cM(s)] > 0$ for $s\in [4m^2,\infty) + i\epsilon$.

Analyticity away from the real axis and the bound $|S(s)|\leq 1$ for $s>4m^2$ imply that $|S(s)|$ is bounded by a constant in the complex $s$ plane, for sufficiently large $|s|$. Hence $|\cM(s)| = O(|s|)$ as $s\rightarrow\infty$ along any direction in the complex plane. We will only need the weaker condition
\be
\frac{\cM(s)}{s^2}\rightarrow 0\quad\text{as}\quad s\rightarrow\infty\,.
\label{eq:MUV}
\ee

\begin{figure}[htb]
\centering
\resizebox{\textwidth}{!}{\pgfdeclaresnake{zigzag}{initial}
{
\state{initial}[width=8pt]
{
\pgfpathlineto{\pgfpoint{2pt}{3pt}}
\pgfpathlineto{\pgfpoint{6pt}{-3pt}}
\pgfpathlineto{\pgfpoint{8pt}{0pt}}
}
\state{final}
{
\pgfpathlineto{\pgfpoint{\pgfsnakeremainingdistance}{0pt}}
}
}

\pgfdeclaresnake{minizigzag}{initial}
{
\state{initial}[width=6pt]
{
\pgfpathlineto{\pgfpoint{1.5pt}{2.25pt}}
\pgfpathlineto{\pgfpoint{4.5pt}{-2.25pt}}
\pgfpathlineto{\pgfpoint{6pt}{0pt}}
}
\state{final}
{
\pgfpathlineto{\pgfpoint{\pgfsnakeremainingdistance}{0pt}}
}
}

\tikzset{cross/.style={cross out, draw=black, minimum size=2*(#1-\pgflinewidth), inner sep=0pt, outer sep=0pt},
  cross/.default={1pt}}

\tikzset{vertpoint/.style={strike out, draw=black, minimum size=2*(#1-\pgflinewidth), inner sep=0pt, outer sep=0pt},
  vertpoint/.default={1pt}}

\begin{tikzpicture}

  \node (0) at (-8, 0) {};
  \node (1) at (-2, 0) {};
  \node (2) at (0, 0) {};
  \node (3) at (6, 0) {};
  \node (4) at (-5.5, -2.25) {};
  \node (5) at (-5.5, 2.35) {};
  \node (6) at (2.5, 2.35) {};
  \node (7) at (2.5, -2.25) {};
  \node (8) at (-3.75, 0) {};
  \node (9) at (4.25, 0) {};
  \node (10) at (3.5, 0) {};
  \node (11) at (2.5, 0) {};
  \node (12) at (1.75, 0) {};
  \node (13) at (-4.5, 0) {};
  \node (14) at (-5.5, 0) {};
  \node (15) at (-6.25, 0) {};
  \node (16) at (-8, 0.225) {};
  \node (17) at (-2, 0.225) {};
  \node (18) at (-8, -0.225) {};
  \node (19) at (-2, -0.225) {};
  \node (20) at (0, 0.225) {};
  \node (21) at (1.75, 0.225) {};
  \node (22) at (4.25, 0.225) {};
  \node (23) at (6, 0.225) {};
  \node (24) at (0, -0.225) {};
  \node (25) at (1.75, -0.225) {};
  \node (26) at (4.25, -0.225) {};
  \node (27) at (6, -0.225) {};
  \node (28) at (-1, 0) {$=$};
  \node (33) at (-2.125, 2.225) {$s$};
  \node (34) at (5.875, 2.225) {$s$};
  \node (35) at (1.875, 0) {};
  \node (36) at (4.125, 0) {};
  \node (37) at (-5, 2.125) {};

  \draw [<-,lightgray] (5.center) to (4.center);
  \draw [<-,lightgray] (6.center) to (7.center);
  \draw [<-,lightgray] (3.center) to (2.center);
  \draw [<-,lightgray] (1.center) to (0.center);
  
  \draw [snake=zigzag,red,thick] (15.center) to (0.center);
  \draw [snake=zigzag,red,thick] (8.center) to (1.center);
  \draw [snake=zigzag,red,thick] (12.center) to (2.center);
  \draw [snake=zigzag,red,thick] (9.center) to (3.center);
  \draw [snake=minizigzag,pink] (14.center) to (15.center);
  \draw [snake=minizigzag,pink] (13.center) to (8.center);
  \draw [snake=minizigzag,pink] (11.center) to (12.center);
  \draw [snake=minizigzag,pink] (10.center) to (9.center);
  
  \draw [blue, bend left=90, looseness=1.1, thick] (16.center) to (17.center);
  \draw [->, blue, thick] (-5.025, 2.155) to (-4.975,2.155);
  \draw [blue, bend right=90, looseness=1.1, thick] (18.center) to (19.center);
  \draw [<-, blue, thick] (-5.025, -2.155) to (-4.975,-2.155);
  \draw [blue, thick] (20.center) to (21.center);
  \draw [->, blue, thick] (0.975,0.225) to (1.025,0.225);
  \draw [blue, thick] (22.center) to (23.center);
  \draw [->, blue, thick] (4.975,0.225) to (5.025,0.225);
  \draw [blue, thick] (24.center) to (25.center);
  \draw [<-, blue, thick] (0.975,-0.225) to (1.025,-0.225);
  \draw [blue, thick] (26.center) to (27.center);
  \draw [<-, blue, thick] (4.975,-0.225) to (5.025,-0.225);
  \draw [blue, in=90, out=90, looseness=1.50, thick] (21.center) to (22.center);
  \draw [->, blue, thick] (2.975,1.3125) to (3.025,1.3125);
  \draw [blue, bend right=90, looseness=1.50, thick] (25.center) to (26.center);
  \draw [<-, blue, thick] (2.975,-1.3125) to (3.025,-1.3125);
  
  \draw [-,black] (33.north west) to (33.south west) to (33.south east);
  \draw [-,black] (34.north west) to (34.south west) to (34.south east);

  \node [cross=4pt, red, thick] at (15.center) {};
  \node [cross=4pt, red, thick] at (8.center) {};
  \node [cross=4pt, red, thick] at (12.center) {};
  \node [cross=4pt, red, thick] at (9.center) {};

  \node [cross=4pt, red, thick] at (14.center) {};
  \node [cross=4pt, red, thick] at (13.center) {};
  \node [cross=4pt, red, thick] at (11.center) {};
  \node [cross=4pt, red, thick] at (10.center) {};
  \node at (-7.5, 2) {\footnotesize{$|s| \rightarrow \infty$}};

  \node at (-6.25, -0.375) {\footnotesize{$4m^2-M^2$}};
  \node at (-4.5, -0.375) {\footnotesize{$4m^2$}};
  \node at (-3.75, -0.375) {\footnotesize{$M^2$}};
  \node at (3.5, -0.475) {\footnotesize{$4m^2$}};
  \node at (4.5, -0.475) {\footnotesize{$M^2$}};
  \node at (1, -0.475) {\footnotesize{$4m^2-M^2$}};

\end{tikzpicture}}
\caption{The contour deformation implementing the dispersive sum rule $\cC_k$ defined in \eqref{eq:CkFlat}.}\label{fig:flatcontour}
\end{figure}
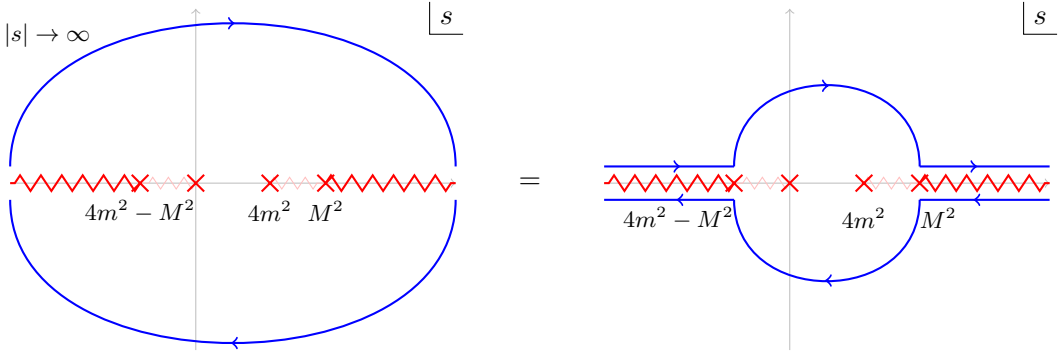

\subsection{Dispersion relations}\label{ssec:DispersionFlat}
Analyticity of $\cM(s)$ away from the real axis, together with the bound \eqref{eq:MUV}, allows one to relate scattering at low and high energy. Concretely, this is achieved through dispersion relations. We obtain an infinite family of dispersive sum rules $\cC_k$ with $k=2,4,6,\ldots$ by noting that, by virtue of the bound \eqref{eq:MUV}, the following contour integral vanishes
\be
\cC_k:\qquad
-\oint\limits_{\infty}\frac{ds}{2\pi i}\frac{1}{s}\frac{\mathcal{M}(s)}{[s(s-4m^2)]^{\frac{k}{2}}}=0\quad\text{for}\quad k=2,4,6,\ldots\,.
\label{eq:CkFlat}
\ee
The contour is an infinitely large circle in the complex $s$-plane, see the left half of Figure \ref{fig:flatcontour}. Next, we choose an energy $M>2m$, and deform the contour as shown in the right half of Figure \ref{fig:flatcontour}. The contour deformation leads to the identity
\be
-\cC_{k}|_{\text{IR}} = \cC_{k}|_{\text{UV}}\,,
\label{eq:IRUV}
\ee
where
\be
-\cC_{k}|_{\text{IR}} = 
\oint\limits_{M^2}\frac{ds}{2\pi i}\frac{1}{s}\frac{\mathcal{M}(s)}{[s(s-4m^2)]^{\frac{k}{2}}}
\label{eq:CIR}
\ee
corresponds to the pair of arcs connecting $s=M^2$ and $s=4m^2-M^2$. The contour on the RHS of \eqref{eq:CIR} runs in the counterclockwise direction. This is the low-energy contribution to the sum rule. On the other hand,
\be
\cC_{k}|_{\text{UV}} =
\int\limits_{M^2}^{\infty}\frac{ds}{\pi}\frac{2(s-2m^2)}{[s(s-4m^2)]^{\frac{k+2}{2}}}\mathrm{Im}[\cM(s)]
\label{eq:CUV}
\ee
comes from the segments connecting $s=M^2$ with $s=+\infty$ and $s=4m^2-M^2$ with $s=-\infty$ along the real axis. We used crossing symmetry of $\cM(s)$ together with $\cM(\overline{s}) = \overline{\cM(s)}$ to combine them to a single integral.

We will refer to the RHS of \eqref{eq:CIR} as the arc at the scale $M$, and denote it by
\be
\cA_k(M) \equiv -\cC_{k}|_{\text{IR}}\,.
\ee
The dispersion relation \eqref{eq:IRUV} states that $\cA_k(M)$ is equal to the moment of $\mathrm{Im}[\cM(s)]$ shown on the RHS of \eqref{eq:CUV}.

The arcs $\cA_k(M)$ satisfy various inequalities as a result of UV unitarity $\mathrm{Im}[\cM(s)]\geq 0$. On the other hand, they can be computed using the low-energy effective field theory (EFT), provided the EFT is valid up to the scale $M$. In this way, dispersion relations lead to interesting bounds on the parameters of the EFT as a consequence of UV consistency.

For concreteness, let us assume that the tree-level approximation in the EFT is valid up to scale $M$. This can be made rigorous by considering a one-parameter family of S-matrices which are becoming arbitrarily weakly coupled. Let us also assume for simplicity that $\phi$ (or $\psi$ in the fermionic case) is the only particle in the EFT and that the theory is invariant under $\phi\mapsto-\phi$, to prevent cubic self-interaction. The most general tree-level amplitude then takes the form
\be
\cM_{\text{EFT}}(s) = \sum\limits_{\substack{k=0\\k\text{ even}}}^{\infty}g_k[s(s-4m^2)]^{\frac{k}{2}}\,,
\label{eq:MEFT}
\ee
where $g_k$ are the EFT couplings, and $g_0 = 0$ in the fermionic case. In the bosonic case, the term $g_k[s(s-4m^2)]^{\frac{k}{2}}$ comes from a contact diagram with four $\phi$s and $2k$ derivatives, i.e. it has the schematic form $g_k(\phi\partial^{k}\phi)^2$ and $k=0,2,4,\ldots$. For fermions, it corresponds to an interaction with four $\psi$s and $2(k-1)$ derivatives, i.e. $g_k(\psi\partial^{k-1}\psi)^2$, and we have $k=2,4,6,\ldots$.\footnote{For this to hold, we should also subtract multiples of lower-derivative contact diagrams to make the amplitude go like $(s-4m^2)^{\frac{k}{2}}$ as $s\rightarrow 4m^2$. In what follows, we parametrize the couplings using their S-matrix elements as in \eqref{eq:MEFT}. We will not need the precise form of the vertices corresponding to $g_k$ appearing in the Lagrangian.}

Assuming the amplitude \eqref{eq:MEFT} provides a reliable approximation to the true amplitude up to $s=M^2$, a residue calculation gives a very simple answer for the arcs
\be
\cA_k(M) = g_k\,.
\label{eq:arcEFT}
\ee
In fact, this was the main motivation for choosing the dispersion relation kernels as in \eqref{eq:CkFlat}. Since $k\geq 2$, the only coupling which is not captured by UV-computable arcs is $g_0$. It is only present in the bosonic case. Additional light matter fields in the EFT would not affect the result \eqref{eq:arcEFT}. This is because the sum of tree-level exchanges
\be
\frac{1}{s-m^2_1}+\frac{1}{t-m^2_1}+\frac{1}{u-m^2_1}
\ee
does not grow as $s\rightarrow\infty$, and thus its $\cA_k$ vanishes, as can be seen by deforming the contour to $s\rightarrow\infty$.

\subsection{Bounds satisfied by the arcs}
Let us discuss bounds on the arcs in more detail. Unless stated otherwise, we do not assume the validity of any specific EFT description, such as \eqref{eq:MEFT}.

First of all, positivity of the integrand in \eqref{eq:CUV} immediately implies
\be
\cA_k(M)>0\quad\text{for all}\quad k=2,4,\ldots \quad\text{and}\quad M>2m\,.
\ee
To obtain more refined bounds, let us perform the change of variables
\be
x = \frac{M^2(M^2-4m^2)}{s(s-4m^2)}
\ee
to arrive at
\be
[M^2(M^2-4m^2)]^{\frac{k}{2}}\cA_k(M) = \int\limits_{0}^{1}dx\,x^{\frac{k-2}{2}}\,\frac{\mathrm{Im}[\cM(s(x))]}{\pi}\,.
\label{eq:arcMoment}
\ee
In other words, $[M^2(M^2-4m^2)]^{\frac{k}{2}}\cA_k(M)$ are moments of a measure in the unit interval. The question of determining a necessary and sufficient condition for a sequence of positive numbers to arise as moments of a measure in a compact interval is known as the Hausdorff moment problem. It has a known solution, reviewed in \cite{Bellazzini:2020cot}. More generally, we can ask for the complete set of constraints on a finite subset of moments. This too has a known solution in terms of Hankel matrices \cite{Arkani-Hamed:2020blm,Bellazzini:2020cot}.

Here, we will confine ourselves to reviewing the complete set of bounds on a pair of arcs $\cA_k(M)$, $\cA_{\ell}(M)$, where $k,\ell$ are positive even integers with $\ell>k$. First, let us introduce the normalized arcs
\be
\widehat{\cA}_k(M)\equiv [M^2(M^2-4m^2)]^{\frac{k-2}{2}}\frac{\cA_k(M)}{\cA_{2}(M)}\,.
\ee
From \eqref{eq:arcMoment}, we have
\be
\widehat{\cA}_k(M) =  \int\limits_{0}^{1}dx\,w(x) x^{\frac{k-2}{2}}\,,\quad\text{where}\quad w(x)\geq 0\quad\text{and}\quad \int\limits_{0}^{1}dx\,w(x) = 1\,. 
\ee
Clearly, $0\leq \widehat{\cA}_k(M) \leq 1$ for all $k$. Furthermore, $\widehat{\cA}_k(M)$ for a fixed $k>2$ can take any value between $0$ and $1$ by setting $w(x)=\delta(x-x_0)$ and varying $x_0$. For a fixed $\widehat{\cA}_k(M)$, $\widehat{\cA}_{\ell}(M)$ must satisfy the lower and upper bounds 
\be
[\widehat{\cA}_{k}(M)]^{\frac{\ell-2}{k-2}}\leq
\widehat{\cA}_{\ell}(M)\leq\widehat{\cA}_{k}(M)\quad\text{for}\quad \ell>k\,.
\label{eq:boundsFlat}
\ee
The upper bound follows from $x^{\frac{\ell-2}{2}}\leq x^{\frac{k-2}{2}}$ for all $x\in[0,1]$. The bound is sharp since it is saturated by the measure $w(x) = \lambda \delta(x) + (1-\lambda)\delta(x-1)$ with varying $\lambda$. On the other hand, the lower bound is saturated by the measure $w(x) = \delta(x-x_0)$ with varying $x_0$. We give two proofs of the lower bound. The first proof uses the H\"older inequality
\be
\int\limits_{0}^{1}dx |f(x)g(x)| \leq \left[\int\limits_{0}^{1}dx |f(x)|^{p}\right]^{\frac{1}{p}}
\left[\int\limits_{0}^{1}dx |g(x)|^{q}\right]^{\frac{1}{q}}\quad\text{where}\quad
p,q\geq 1\,,\quad \frac{1}{p}+\frac{1}{q} = 1\,.
\ee
The lower bound follows immediately after setting
\be
f(x) = w(x)^{\frac{\ell-k}{\ell-2}}\,,\quad
g(x) = w(x)^{\frac{k-2}{\ell-2}}x^{\frac{k-2}{2}}\,,\quad
p = \frac{\ell-2}{\ell-k}\,,\quad
q = \frac{\ell-2}{k-2}\,.
\ee

The second proof of the lower bound is a direct application of Jensen's inequality. Let $w(x)$ be a probability measure in the unit interval as above, let $f(x)$ be a convex function and let $g(x)$ be any (measurable) real-valued function. Jensen's inequality states
\be
f\!\left(\int\limits_{0}^{1}dx\, w(x) g(x) \right) \leq \int\limits_{0}^{1}dx\, w(x) f(g(x))\,.
\label{eq:jensen}
\ee
The lower bound in \eqref{eq:boundsFlat} follows from this after setting
\be
f(x) = x^{\frac{\ell-2}{k-2}}\,,\quad g(x) = x^{\frac{k-2}{2}}\,.
\ee
Since $\ell>k$, $f(x)$ is indeed convex as required. It is the second proof which will generalize to the AdS case.

When the tree-level EFT description \eqref{eq:MEFT} is valid, the above bounds translate into bounds on the four-point couplings $g_k$. We get the strongest bounds by choosing the largest possible $M$ where \eqref{eq:MEFT} converges. Thus $M$ is the mass of the lightest state not included in the EFT. Firstly, we have $g_k\geq 0$ for all $k\geq 2$. The bound $\widehat{\cA}_{\ell}(M) \leq \widehat{\cA}_{k}(M)$ for all $2\leq k\leq \ell$ becomes
\be
0\leq \frac{g_{\ell}}{g_k} \leq \frac{1}{[M^2(M^2-4m^2)]^{\frac{\ell-k}{2}}}\quad\text{for}\quad
2\leq k\leq \ell\,.
\label{eq:upperFlat}
\ee
Thus, these ratios are bounded by inverse powers of the UV cutoff dictated by the dimensional analysis. On the other hand, the lower bound $[\widehat{\cA}_{k}(M)]^{\frac{\ell-2}{k-2}}\leq\widehat{\cA}_{\ell}(M)$ for $\ell\geq k$ becomes
\be
\left(\frac{g_{\ell}}{g_2}\right)^{k-2}\geq \left(\frac{g_k}{g_2}\right)^{\ell-2}
\quad\text{for}\quad
2\leq k\leq \ell\,,
\label{eq:lowerFlat}
\ee
with no $M$-dependence. These inequalities can be tested by performing low-energy measurements without knowing the UV cutoff $M$.

\subsection{Bounds at one loop}
We will conclude this section by a discussion of how the bounds are corrected by one-loop effects in the EFT. For simplicity, we will focus on scattering of massless particles, i.e.\ $m=0$, and suppose that they are derivatively coupled, i.e.\ $g_0 = 0$. This is automatic in the fermionic case. In the bosonic case, this set-up applies to scattering of excitations on a flux tube in three dimensions. The flux-tube theory is weakly-coupled in the limit of a large number of colors. Bootstrap of the flux-tube S-matrix at finite coupling was performed in the interesting recent papers \cite{EliasMiro:2019kyf,EliasMiro:2021nul}.

At one loop, the tree-level amplitude \eqref{eq:MEFT} gets corrected to

\begin{equation}
  \mathcal{M}_{\text{EFT}}(s) = \sum\limits_{\substack{k=2\\k\text{ even}}}^{\infty}g_ks^k +\frac{i}{4} \sum\limits_{\substack{k,\ell=2\\k,\ell\text{ even}}}^{\infty}g_kg_{\ell}s^{k+\ell-1}
  \quad\text{for}\quad \mathrm{Im}(s)>0\,.
\end{equation}
This can be seen most quickly from \eqref{eq:UnitOnM} by noting that at one loop in the EFT, the inequality becomes an equality. The expression in the lower half-plane follows from $\mathcal{M}(\overline{s})=\overline{\mathcal{M}(s)}$. The upper and lower half-planes are seperated by a branch cut that now extends along the entire real axis

The one-loop effects modify the expression for the arcs in terms of the low-energy couplings
\begin{equation}
  \label{eq:sum-rule-loop}
  \mathcal{A}_k(M) = g_k - \frac{1}{2\pi}\sum\limits_{\substack{i,j=2\\ i,j\text{ even}}}^\infty\frac{g_ig_j}{i+j-k-1}M^{2(i+j-k-1)}.
\end{equation}

\begin{figure}[h]
  \centering
  \includegraphics[width=0.5\textwidth]{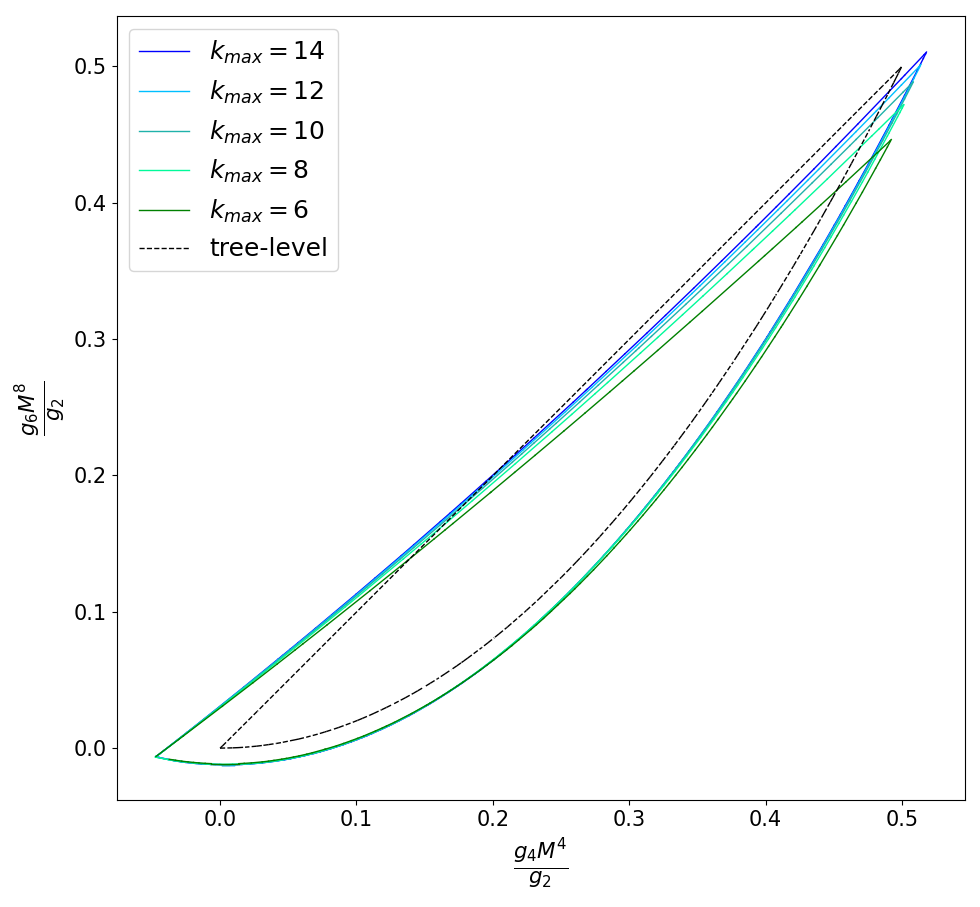}
  \caption{Exclusion plot of $g_6$ against $g_4$  for $g_2=0.5$ including couplings up to $g_{k_{\text{max}}}$, tree-level bound for reference. The allowed region is within the banana shape.}
  \label{fig:0.5-bound}
\end{figure}

\begin{figure}[h]
  \centering
  \includegraphics[width=0.5\textwidth]{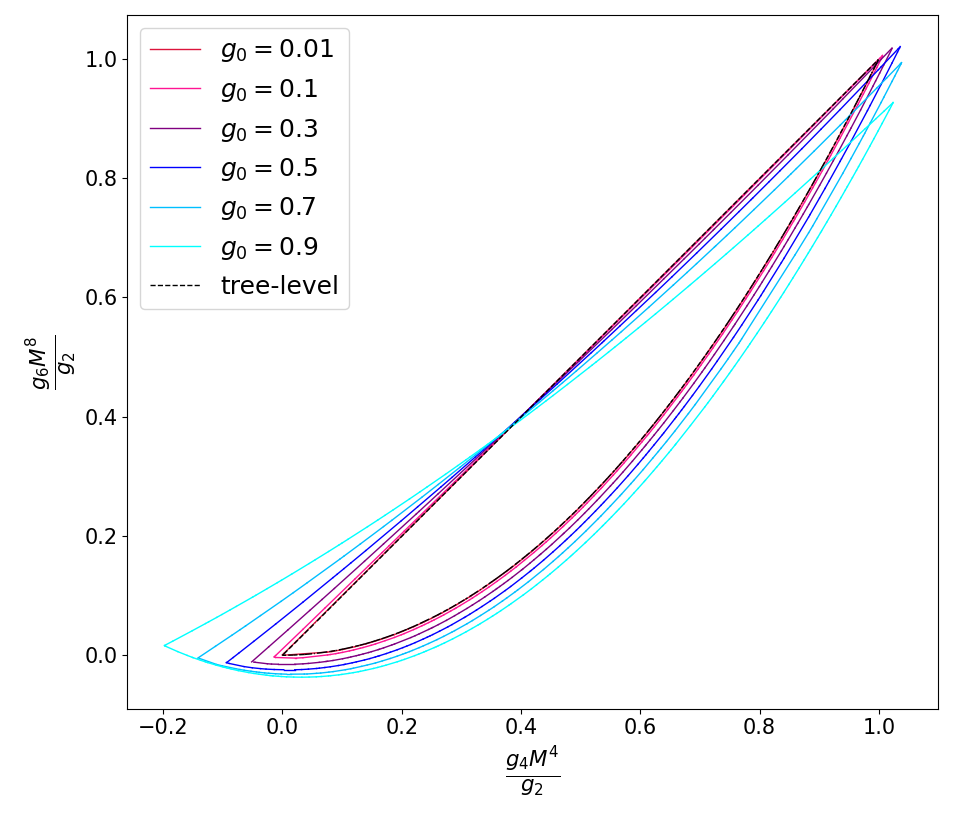}
  \caption{Exclusion plot of $g_6$ against $g_4$ for various values of $g_2$, tree-level bound for reference. To produce the bounds, we set $g_k = 0$ for all $k>14$.}
  \label{fig:change-g0-bound}
\end{figure}

We would like to analyze the effect of this correction on the bounds in the space of the $g_{k}$ couplings. Let us focus on the first three couplings $g_2,\,g_4,\,g_6$. Using the UV cutoff $M$, we can form the dimensionless combinations
\be
y_4 = \frac{M^4 g_{4}}{g_2}\qquad y_6 = \frac{M^8 g_{6}}{g_2}\,.
\ee
The tree-level bounds on $y_4$, $y_6$ derived in the previous subsection are $0\leq y_4 \leq 1$, $0\leq y_6 \leq 1$, together with
\be
y^2_4\leq y_6 \leq y_4\,.
\ee
To find the one-loop corrections to these bouds, we can use the following inequalities satisfied by the arcs $\mathcal{A}_{2}$, $\mathcal{A}_{4}$ and $\mathcal{A}_{6}$

\begin{equation}
    M^4\mathcal{A}_4-\mathcal{A}_6 \geq 0\,, \qquad \mathcal{A}_6\mathcal{A}_2 - \mathcal{A}_4^2 \geq 0
    \label{eq:arcs46}
\end{equation}
The one-loop bounds are shown in Figures \ref{fig:0.5-bound} and \ref{fig:change-g0-bound}. To produce the plots, we truncate the expansion of the arcs \eqref{eq:sum-rule-loop} to finitely many couplings $g_k$ with $k=2,4,\ldots,k_{\text{max}}$ and ask if the inequalities \eqref{eq:arcs46} are satisfied. Note that since the one-loop arcs are not homogeneous in the couplings, the bounds have a nontrivial dependence on the value of $g_2$.

As expected, for very small $g_2$, the region differs very little from the tree-level result. The one-loop corrections weaken the tree-level bounds. Importantly, they now allow for slightly negative $g_4$ and $g_6$. A lower bound on $g_4$ for $g_2$ was found in \cite{EliasMiro:2019kyf,EliasMiro:2021nul} using non-perturbative methods. This gives $g_4 \geq -\frac{1}{768}$, which is saturated by the following S-matrix: $S(s) = \frac{i8-s}{i8+s}$. The one-loop bound found here includes this point, in contrast with the tree-level bound.

\section{Scattering in AdS$_2$}\label{sec:CFT}
\subsection{1d CFT kinematics}
We now wish to study the same scattering processes in AdS space. We imagine placing the QFT in $\mathrm{AdS}_2$ of radius $R$. This defines a consistent 1d CFT \cite{Paulos:2016fap}. By a 1d CFT we mean the following. The space of states is a unitary representation of the Lorentzian conformal group $\widetilde{SO}(2,1)$. Each of its irreducible components gives rise to a primary operator which can be inserted on the AdS boundary. These operators admit a standard operator product expansion and their correlators satisfy the constraints of conformal invariance, unitarity and crossing symmetry. Compared to gravitational holography, the main difference is the absence of a boundary stress tensor. However, this distinction does not preclude a successful application of the conformal bootstrap to these theories.\footnote{Reference \cite{Paulos:2016fap} reserves the term conformal \emph{field} theory to theories with a stress tensor, and refers to the present case without a stress tensor as conformal theory.}

Let us vary the AdS radius $R$ while keeping the masses and couplings of the bulk theory fixed. We obtain a continuous 1-parameter family of 1d CFTs. The flat-space theory is encoded in the large-$R$ properties of this family. If the flat space theory contains a particle of mass $m$, the CFT contains a primary operator of dimension $\Delta>0$, satisfying
\be
\Delta(\Delta-1)= m^2R^2\,.
\ee
Thus, $\Delta\sim m R$ as $R\rightarrow\infty$.

To compare with our flat-space analysis, we will study scattering of identical particles of mass $m$ in $\mathrm{AdS}_2$. When the scattered particles are bosons, we will denote their dual primary operator $\phi$, and $\psi$ when they are fermions. Two-to-two scattering is encoded in the four-point functions $\langle\phi\phi\phi\phi\rangle$, $\langle\psi\psi\psi\psi\rangle$. 

We will first discuss the four-point functions in Euclidean kinematics. This means that we consider the bulk theory in the upper half-plane, parametrized by coordinates $x,y\in\mathbb{R}$ with $y>0$ and endowed with the metric
\be
ds^2 = R^2 \frac{dx^2+dy^2}{y^2}\,.
\ee
The operators of the dual 1d CFT live on the boundary $y=0$ and are parametrized by the Euclidean coordinate $x$. Due to conformal invariance, their four-point functions take the form
\ba
\langle\phi(x_1)\phi(x_2)\phi(x_3)\phi(x_4)\rangle &= \langle\phi(x_1)\phi(x_3)\rangle\langle\phi(x_2)\phi(x_4)\rangle\cG_{\rB}(z)\\
\langle\psi(x_1)\psi(x_2)\psi(x_3)\psi(x_4)\rangle &=
\langle\psi(x_1)\psi(x_3)\rangle\langle\psi(x_2)\psi(x_4)\rangle\cG_{\rF}(z)\,.
\ea
Here and in the following, the B,F label stands for boson, fermion. We have
\be
\langle\phi(x_i)\phi(x_j)\rangle = \frac{1}{|x_{ij}|^{2\Df}}\,,\qquad
\langle\psi(x_i)\psi(x_j)\rangle = \frac{\mathrm{sgn}(x_{ij})}{|x_{ij}|^{2\Dp}}\,,
\ee
where $x_{ij} = x_i-x_j$, and
\be
z = \frac{x_{12}x_{34}}{x_{13}x_{24}}
\ee
is the cross-ratio. Without loss of generality, we can focus on the ordering $x_1<x_2<x_3<x_4$, which corresponds to $0<z<1$. Permutation symmetry under $x_2\leftrightarrow x_4$ leads to crossing symmetry
\be
\cG_{\rB}(z) = \cG_{\rB}(1-z)\,,\qquad
\cG_{\rF}(z) = \cG_{\rF}(1-z)
\label{eq:crossing}
\ee
for all $z\in(0,1)$. Note that there is no minus sign in the fermionic case.

The four-point function can be expanded using the s-channel OPE
\be
\cG_{\rB}(z) = \sum\limits_{\cO} (c_{\phi\phi\cO})^2 G^{\Df}_{\Delta_{\cO}}(z)\,,\qquad
\cG_{\rF}(z) = \sum\limits_{\cO} (c_{\psi\psi\cO})^2 G^{\Dp}_{\Delta_{\cO}}(z)\,,
\label{eq:OPE}
\ee
where the sums run over primary operators in the respective OPEs, $c_{\phi\phi\cO},c_{\psi\psi\cO}\in\mathbb{R}$ are the OPE coefficients and
\be
G^{\Df}_{\Delta}(z) = z^{\Delta-2\Df}{}_2F_1(\Delta,\Delta;2\Delta,z)
\ee
are the $sl_2(\mathbb{R})$ conformal blocks. A standard argument using the $\rho$-coordinate \cite{Hogervorst:2013sma} and positivity of $(c_{\phi\phi\cO})^2,(c_{\psi\psi\cO})^2$ shows that the s-channel OPE \eqref{eq:OPE} converges away from the t-channel branch cut $z\in[1,\infty)$, and that the analytic continuation of $\cG_{\rB}(z),\,\cG_{\rF}(z)$ to the cut plane $\mathbb{C}\backslash (-\infty,0]\cup[1,\infty)$ is holomorphic.

Combining crossing symmetry with the OPE leads to the conformal bootstrap equations
\ba
&\sum\limits_{\cO} (c_{\phi\phi\cO})^2\left[G^{\Df}_{\Delta_{\cO}}(z) - G^{\Df}_{\Delta_{\cO}}(1-z)\right] = 0\\
&\sum\limits_{\cO} (c_{\psi\psi\cO})^2\left[G^{\Dp}_{\Delta_{\cO}}(z) - G^{\Dp}_{\Delta_{\cO}}(1-z)\right] = 0
\,.
\label{eq:bootstrapEq}
\ea
These equations hold in the Euclidean region $0<z<1$, as well as everywhere in the cut plane $z\in \mathbb{C}\backslash (-\infty,0]\cup[1,\infty)$. As we will review in the following subsection, the analytic continuation of $\cG(z)$ to $z<0$ corresponds to Lorentzian kinematics. Studying the crossing equation in this kinematics will allow us to derive the AdS analogue of dispersive sum rules of Section \ref{sec:SMatrix}.

\subsection{The double commutator}
The defining property of flat-space dispersive sum rules is that the heavy states enter through the discontinuity of the amplitude. It is well-understood that the notion which plays the role of the discontinuity in the context of AdS scattering is the double commutator of the boundary correlation function \cite{Caron-Huot:2017vep}.

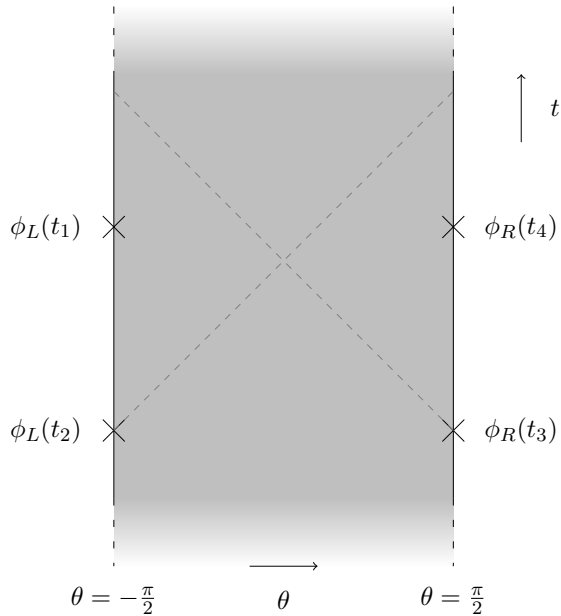
\begin{figure}[htb]
\centering
\resizebox{0.5\textwidth}{!}{\tikzset{cross/.style={cross out, draw=black, minimum size=2*(#1-\pgflinewidth), inner sep=0pt, outer sep=0pt},
cross/.default={1pt}}

\begin{tikzpicture}\label{fig:globalAdS}
  \node (0) at (-3, 7.25) {};
  \node (1) at (-3, 1) {};
  \node (2) at (2, 7.25) {};
  \node (3) at (2, 1) {};
  \node (4) at (-3, 8.25) {};
  \node (5) at (2, 8.25) {};
  \node (6) at (-3, 0) {};
  \node (7) at (2, 0) {};
  \node (8) at (-3, 2) {};
  \node (9) at (2, 2) {};
  \node (10) at (2, 7) {};
  \node (11) at (-3, 7) {};
  \node (12) at (-3, 5) {};
  \node (13) at (2, 5) {};
  \node (14) at (3, 7.25) {};
  \node (15) at (3, 6.25) {};
  \node (16) at (-1, 0) {};
  \node (17) at (0, 0) {};
  \node (18) at (-0.5, -0.5) {};
  \node (19) at (-0.5, -0.5) {$\theta$};
  \node (20) at (3.5, 8.5) {};
  \node (21) at (3.5, 6.75) {$t$};
  \node (22) at (-4, 5) {};
  \node (23) at (-4, 5) {$\phi_L (t_1)$};
  \node (24) at (3, 5) {};
  \node (25) at (3, 5) {$\phi_R (t_4)$};
  \node (26) at (-4, 2) {$\phi_L (t_2)$};
  \node (27) at (3, 2) {};
  \node (28) at (3, 2) {$\phi_R (t_3)$};
  \node (29) at (-3, -0.5) {};
  \node (30) at (-3, -0.5) {$\theta = -\frac{\pi}{2}$};
  \node (31) at (2, -0.5) {$\theta = \frac{\pi}{2}$};

  \fill [lightgray] (0.center) to (1.center) to (3.center) to (2.center) to (0.center);
  \fill [lightgray, path fading = north] (0.center) to (4.center) to (5.center) to (2.center) to (0.center);
  \draw [lightgray] (0.center) to (2.center);
  \fill [lightgray, path fading = south] (1.center) to (6.center) to (7.center) to (3.center) to (1.center);
  \draw [lightgray] (1.center) to (3.center);
  
  \node [cross=5pt] (32) at (-3, 2) {};
  \node [cross=5pt] (33) at (2, 2) {};
  \node [cross=5pt] (34) at (2, 5) {};
  \node [cross=5pt] (35) at (-3, 5) {};
  
  \draw (0.center) to (1.center);
  \draw (2.center) to (3.center);
  \draw [style=loosely dashed] (5.center) to (2.center);
  \draw [style=loosely dashed] (4.center) to (0.center);
  \draw [style=loosely dashed] (1.center) to (6.center);
  \draw [style=loosely dashed] (3.center) to (7.center);
  \draw [style=dashed, color=gray] (8.center) to (10.center);
  \draw [style=dashed, color=gray] (9.center) to (11.center);
  
  \draw [<-] (14.center) to (15.center);
  \draw [<-] (17.center) to (16.center);
  
\end{tikzpicture}}
\caption{Kinematics in which we define the double commutator on the boundary of AdS$_2$.}\label{fig:globalAdS2}
\end{figure}

To construct the double commutator on the boundary of AdS$_2$, we can consider Lorentzian AdS$_2$ in global coordinates $(t,\theta)$, see Figure \ref{fig:globalAdS2}. Here $t\in \mathbb{R}$ is the global Lorentzian time and $\theta\in(-\pi/2,\pi/2)$ is the spatial coordinate. The metric takes the form
\be
ds^2 = R^2 \frac{-dt^2+d\theta^2}{\cos^2\theta}\,.
\ee
Thus, global AdS$_2$ is an infinite strip with timelike boundaries at $\theta = \pm \pi/2$. Let us consider the double commutator with a pair of operators on each boundary
\be
\langle \Omega|
[\phi_{\text{R}}(t_3),\phi_{\text{R}}(t_4)]
[\phi_{\text{L}}(t_1),\phi_{\text{L}}(t_2)]
|\Omega\rangle\,.
\label{eq:DC}
\ee

Here $\phi_{\text{L}}(t_1),\,\phi_{\text{L}}(t_2)$ are inserted at $\theta = -\pi/2$ and $\phi_{\text{R}}(t_3),\,\phi_{\text{R}}(t_4)$ at $\theta = \pi/2$. Let us assume $t_1>t_2$, $t_4>t_3$, $t_1-t_3<\pi$, $t_4-t_2<\pi$, so that all pairs except for 12 and 34 are spacelike separated through the interior of AdS$_2$. The analytic continuation of the Euclidean four-point function $\cG_\rB(z)$ encodes arbitrary Wightman four-point functions. To evaluate \eqref{eq:DC} in terms of $\cG_\rB(z)$, we can transform back to the Euclidean coordinates
\be
x = e^{\tau}\sin\theta\,,\quad y = e^{\tau}\cos\theta\,,
\ee
where $\tau = i t$ is the global Euclidean time. After some care with the $i\epsilon$ prescription, we find

\ba
\frac{1}{2}\langle \Omega|&
[\phi_{\text{L}}(t_1),\phi_{\text{L}}(t_2)]
[\phi_{\text{R}}(t_3),\phi_{\text{R}}(t_4)]
|\Omega\rangle = \\
&=\langle \Omega|\phi_{\text{L}}(t_1)\phi_{\text{R}}(t_3)|\Omega\rangle\,
\langle \Omega|\phi_{\text{L}}(t_2)\phi_{\text{R}}(t_4)|\Omega\rangle\,
\dDisc_{\rB}\cG_{\rB}(z)\,,
\ea
where $z\in(-\infty,0)$ and the double discontinuity is defined as follows
\be
\dDisc_{\rB}\cG_{\rB}(z) = -
\tfrac{1}{2}\cG^{\curvearrowleft}_{\rB}(z)-
\tfrac{1}{2}\cG^\text{\rotatebox[origin=c]{180}{\reflectbox{$\curvearrowleft$}}}_{\rB}(z)
+(1-z)^{-2\Df}\cG_{\rB}\!\left(\tfrac{z}{z-1}\right)\,.
\ee
The arrows show the path of analytic continuation from $z\in(0,1)$ to $z<0$. For completeness, let us record the cross-ratio in terms of the global times
\be
z = \frac{\sin\left(\tfrac{t_{12}}{2}\right)\sin\left(\tfrac{t_{34}}{2}\right)}{\cos\left(\tfrac{t_{13}}{2}\right)\cos\left(\tfrac{t_{24}}{2}\right)}\in(-\infty,0)\,.
\ee
In the fermionic case, the double commutator is replaced by the double anti-commutator, leading to
\ba
\frac{1}{2}\langle \Omega|&
\{\psi_{\text{L}}(t_1),\psi_{\text{L}}(t_2)\}
\{\psi_{\text{R}}(t_3),\psi_{\text{R}}(t_4)\}
|\Omega\rangle = \\
&=\langle \Omega|\psi_{\text{L}}(t_1)\psi_{\text{R}}(t_3)|\Omega\rangle\,
\langle \Omega|\psi_{\text{L}}(t_2)\psi_{\text{R}}(t_4)|\Omega\rangle\,
\dDisc_{\rF}\cG_{\rF}(z)\,,
\ea
where
\be
\dDisc_{\rF}\cG_{\rF}(z) = 
\tfrac{1}{2}\cG^{\curvearrowleft}_{\rF}(z)+
\tfrac{1}{2}\cG^\text{\rotatebox[origin=c]{180}{\reflectbox{$\curvearrowleft$}}}_{\rF}(z)
+(1-z)^{-2\Dp}\cG_{\rF}\!\left(\tfrac{z}{z-1}\right)\,.
\ee
The double discontinuity of a correlator can be evaluated using the s-channel OPE. Its action on individual conformal block is
\ba
\dDisc_{\text{B}}G^{\Df}_{\Delta}(z) &= 2\sin^2\!\left[\tfrac{\pi}{2}(\Delta-2\Df)\right]\widehat{G}^{\Df}_{\Delta}(z)\\
\dDisc_{\text{F}}G^{\Dp}_{\Delta}(z) &= 2\cos^2\!\left[\tfrac{\pi}{2}(\Delta-2\Dp)\right]\widehat{G}^{\Dp}_{\Delta}(z)\,,
\ea
where
\be
\widehat{G}^{\Df}_{\Delta}(z) = (-z)^{\Delta-2\Df}{}_2F_1(\Delta,\Delta;2\Delta,z)\,.
\label{eq:gHat}
\ee
Crucially, dDisc exhibits double zeros on the double-trace dimensions of mean field theory, namely $\Delta^{\text{B}}_n = 2\Df+2n$ for bosons and  $\Delta^{\text{F}}_n = 2\Dp+2n+1$ for fermions.

\subsection{Dispersive sum rules}\label{ssec:disperiveReview}
Dispersive sum rules arise by applying suitable linear functionals to the conformal bootstrap equation \eqref{eq:bootstrapEq}. The functionals are constructed so that all sufficiently heavy operators enter only through their dDisc. In $d>1$ CFTs, this can be achieved by integrating the Lorentzian crossing equation $\langle\Omega|\phi(x_4)[\phi(x_3),\phi(x_1)]\phi(x_2)|\Omega\rangle = 0$ with respect to $x_1$ and $x_3$ along spacelike separated null rays \cite{Kologlu:2019bco,Caron-Huot:2020adz,Caron-Huot:2021enk}. While there are no null rays in 1d spacetime, it is nevertheless possible to construct dispersive functionals for 1d CFTs by integrating \eqref{eq:bootstrapEq} in the complex $z$ plane against judiciously chosen kernels \cite{Mazac:2016qev,Mazac:2018mdx,Mazac:2018ycv}. Let us quickly review the construction. We refer the reader to \cite{Mazac:2018mdx} for details.

Consider a linear functional $\omega$ acting on functions $\cG(z)$ in the cut plane. Suppose $\omega$ is antisymmetric under crossing, i.e. $\omega[\cG(z)] = -\omega[\cG(1-z)]$ for all $\cG(z)$.\footnote{For example, $\omega$ can be a linear combination of derivatives of odd order at $z=1/2$, which is the standard choice in the numerical conformal bootstrap.} It follows that if $\cG(z) = \cG(1-z)$, we must have $\omega[\cG] = 0$. Provided $\omega$ commutes with the sum over conformal blocks, we get the sum rule\footnote{From now on, we will treat the bosonic and fermionic cases in parallel, but will only give explicit formulas for bosons, unless the fermionic case is different.}
\be
\sum\limits_{\cO} (c_{\phi\phi\cO})^2 \omega[G^{\Df}_{\Delta_{\cO}}] = 0\,.
\label{eq:sumRule}
\ee

\begin{figure}[ht]
\centering
\resizebox{0.6\textwidth}{!}{\pgfdeclaresnake{zigzag}{initial}
{
\state{initial}[width=4pt]
{
\pgfpathlineto{\pgfpoint{1pt}{2pt}}
\pgfpathlineto{\pgfpoint{3pt}{-2pt}}
\pgfpathlineto{\pgfpoint{4pt}{0pt}}
}
\state{final}
{
\pgfpathlineto{\pgfpoint{\pgfsnakeremainingdistance}{0pt}}
}
}

\pgfdeclaresnake{inversezigzag}{initial}
{
\state{initial}[width=4pt]
{
\pgfpathlineto{\pgfpoint{1pt}{-2pt}}
\pgfpathlineto{\pgfpoint{3pt}{2pt}}
\pgfpathlineto{\pgfpoint{4pt}{0pt}}
}
\state{final}
{
\pgfpathlineto{\pgfpoint{\pgfsnakeremainingdistance}{0pt}}
}
}

\tikzset{vertpoint/.style={strike out, draw=black, minimum size=2*(#1-\pgflinewidth), inner sep=0pt, outer sep=0pt},
vertpoint/.default={1pt}}

\tikzset{cross/.style={cross out, draw=black, minimum size=2*(#1-\pgflinewidth), inner sep=0pt, outer sep=0pt},
cross/.default={1pt}}

\begin{tikzpicture}
  \fill [lightgray] (-5, 3) to (5, 3) to (5,-3) to (-5,-3) to (-5, 3);
  
  \node (3) at (0, -1.5) {};
  \node (4) at (0, -3) {};
  \node (5) at (0, 3) {};
  \node (6) at (-1, 0) {};
  \node (7) at (-5, -3) {};
  \node (8) at (5, -3) {};
  \node (9) at (5, 4) {};
  \node (10) at (-2, 0) {};
  \node (11) at (0, 0) {};
  \node (12) at (2, 0) {};
  \node (13) at (-5, 0) {};
  \node (14) at (5, 0) {};
  \node (15) at (0, 1.5) {};
  \node (16) at (1, 0) {};

  \draw [-, gray] (13.center) to (14.center);
  
  \draw [snake=zigzag, red] (13.center) to (10.center);
  \draw [snake=zigzag, red] (12.center) to (14.center);
  
  \draw [->, very thick] (11.center) to (15.center);
  \draw [-, very thick] (15.south) to (5.center);
  \node at (0.5, 2) {$f(z)$};
  \draw [->, very thick] (11.center) to (16.center);
  \draw [-, very thick] (16.west) to (12.center);
  \node at (1, 0.5) {$g(z)$};

  \draw [->, very thick] (11.center) to (3.center);
  \draw [-, very thick] (3.north) to (4.center);
  \node at (-0.5, -2) {$f(z)$};
  \draw [->, very thick] (11.center) to (6.center);
  \draw [-, very thick] (6.east) to (10.center);
  \node at (-1, 0.5) {$g(1-z)$};

  \node [vertpoint=5pt, black, rotate=45, ultra thick] (1) at (2, 0) {};
  \node at (2, -0.5) {$z=1$};
  \node [vertpoint=5pt, black, rotate=45, ultra thick] (2) at (-2, 0) {};
  \node at (-2, -0.5) {$z=0$};

  \node [cross=1pt, gray] at (11.center) {};
\end{tikzpicture}}
\caption{The contours defining the general dispersive functional \eqref{eq:omegaGeneral}}\label{fig:LContour}
\end{figure}
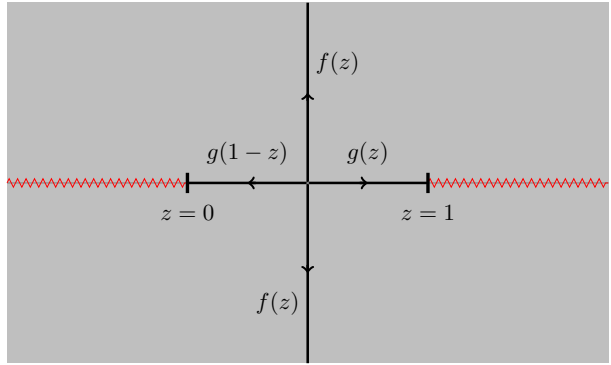

We say that $\omega$ is dispersive if $\omega[G^{\Df}_{\Delta}]$ has a double zero on double trace dimentions $\Delta = 2\Df+2n$ for sufficiently large $n$, i.e. all $n\geq n_*$. Reference \cite{Mazac:2018mdx} gave a general construction of $\omega$ such that it commutes with the OPE and such that $\omega$ is dispersive. The general form of a dispersive functional is
\be
\omega[\cG] =
\pm\frac{1}{2}\!\!\!\int\limits_{\frac{1}{2}}^{\frac{1}{2}+i\infty}\!\!\!dz f(z)\cG(z)
\pm\frac{1}{2}\!\!\!\int\limits_{\frac{1}{2}}^{\frac{1}{2}-i\infty}\!\!\!dz f(z)\cG(z)+
\int\limits_{\frac{1}{2}}^{1}dz \,g\!\left(z\right)\cG(z)+\int\limits_{\frac{1}{2}}^{0}dz \,g\!\left(1-z\right)\cG(z)\,.
\label{eq:omegaGeneral}
\ee
The contours are shown in Figure \ref{fig:LContour}. The top and bottom sign applies respectively to bosons and fermions. $f(z)$, $g(z)$ are a pair of kernels related as follows
\ba
g(z) =
\begin{cases}
&(1-z)^{2\Df-2}f\left(\tfrac{z}{z-1}\right)\quad\text{(bosons)}\vspace{10pt}\\
&(1-z)^{2\Dp-2}f\left(\tfrac{z}{z-1}\right)\quad\text{(fermions)}\,.
\end{cases}
\label{eq:gFromf}
\ea
$f(z)$ is holomorphic away from a branch cut at $z\in[0,1]$ and satisfies the symmetry
\be
f(z) = f(1-z)
\label{eq:fSymmetry}
\ee
as well as the identity
\ba
&z^{2\Df-2}f\left(\tfrac{1}{z}\right)+(1-z)^{2\Df-2}f\left(\tfrac{1}{1-z}\right)+\mathrm{Re}[f(z)] = 0\quad\text{(bosons)}\\
&z^{2\Dp-2}f\left(\tfrac{1}{z}\right)+(1-z)^{2\Dp-2}f\left(\tfrac{1}{1-z}\right)-\mathrm{Re}[f(z)] = 0\quad\text{(fermions)}
\label{eq:functionalEq}
\ea
for $z\in(0,1)$. The logic behind \eqref{eq:omegaGeneral} and properties \eqref{eq:gFromf}, \eqref{eq:fSymmetry} and \eqref{eq:functionalEq} is the following. Firstly, symmetry \eqref{eq:fSymmetry} guarantees that the functional is antisymmetric under crossing, i.e. $\omega[\cG(z)] = -\omega[\cG(1-z)]$ as required. Secondly, the identities \eqref{eq:gFromf} and \eqref{eq:functionalEq} guarantee that the action of $\omega$ on a general function $\cG(z)$ can be written as an integral of $\dDisc\,\cG$ weighted by $f(z)$
\be
\omega[\cG] = \int\limits_{-\infty}^{0}\!\! dz f(z)\, \dDisc\,\cG(z)\,.
\label{eq:dispersiveF}
\ee
This can be shown by deforming the vertical contours in \eqref{eq:omegaGeneral} to the left-hand branch cut. In particular, the action on the conformal blocks is 
\be
\omega[G^{\Df}_{\Delta}] = 
\begin{cases}
2\sin^2\!\left[\tfrac{\pi}{2}(\Delta-2\Df)\right]\int\limits_{-\infty}^{0}dz f(z)\,\widehat{G}^{\Df}_{\Delta}(z)\quad\text{(bosons)}\vspace{5pt}\\
2\cos^2\!\left[\tfrac{\pi}{2}(\Delta-2\Dp)\right]\int\limits_{-\infty}^{0}dz f(z)\,\widehat{G}^{\Dp}_{\Delta}(z)\quad\text{(fermions)}\,,
\end{cases}
\label{eq:dispersiveManifest}
\ee
where $\widehat{G}^{\Df}_{\Delta}$ is a transformed conformal block defined in \eqref{eq:gHat}. The integral converges for all $\Delta>\Delta_{*}$, manifesting the fact that $\omega$ is dispersive. The value of $\Delta_{*}$ depends on the strength of the singularity of $f(z)$ at $z=0$.

\subsection{The Regge moments}
An important quantity attached to a dispersive functional is its Regge spin $k$. We say that the functional \eqref{eq:omegaGeneral} has Regge spin $k$ if $f(z)\sim b\,z^{-k}$ as $z\rightarrow i\infty$ for some $b\neq 0$. Similarly, we say that a four-point function $\cG(z)$ has Regge spin $J$ if
\be
\cG(z) \sim b\,z^{J-1}\quad\text{as}\quad z\rightarrow i\infty
\ee
for some $b\neq 0$. The reason for this terminology is that the limit $z\rightarrow i\infty$ is analogous to the u-channel Regge limit of four-point functions in $d>1$ CFTs. When $\cG(z)$ arises from a $d>1$ correlator by restricting the operators to lie on a line, then $z\rightarrow i\infty$ is literally the u-channel Regge limit. In that case, the notion of the Regge spin defined here agrees with the standard notion of Regge spin in $d>1$.

Let $\omega$ be a functional of Regge spin $k$ and $\cG(z)$ a crossing-symmetric four-point function of Regge spin $J$. If $k>J$, then $\omega$ can be swapped with the conformal block expansion of $\cG(z)$ and we get the sum rule \eqref{eq:sumRule}. Nonperturbative four-point functions in unitary theories satisfy $J\leq 1$, so only functionals of Regge spin $k>1$ generally lead to valid sum rules. Note that since $f(z)=f(1-z)$, the Regge spin of the dispersive functional \eqref{eq:omegaGeneral} is always an even integer. Requiring the swapping property with physical correlators then restricts the possible values of $k$ to positive even integers.

The Regge spin of a dispersive functional translates to a specific power-law suppression of its action on heavy conformal blocks. Let $\omega$ be a functional of the form \eqref{eq:omegaGeneral} such that $f(z)\sim b\cdot z^{-k}$ as $z\rightarrow i\infty$ for some $b\neq 0$. The leading asymptotics of its action reads
\be
\frac{p^{\text{MFT}}_{\Delta}\,\omega[G^{\Df}_{\Delta}]}{2\sin^2\!\left[\tfrac{\pi}{2}(\Delta-2\Df)\right]}\sim 
 b\cdot \frac{2\,\Gamma (2\Df+k-1)^2}{\Gamma (2 \Df)^2}\Delta^{1-2 k}
 \quad\text{as}\quad \Delta\rightarrow\infty\,.
 \label{eq:heavyOmega}
\ee
We multiplied the action by $p^{\text{MFT}}_{\Delta}$, which is the density of squared OPE coefficients in mean field theory
\be
p^{\text{MFT}}_{\Delta} = \frac{\Gamma (\Delta )^2 \Gamma (\Delta +2 \Df -1)}{\Gamma (2 \Delta -1) \Gamma (2 \Df )^2 \Gamma (\Delta -2 \Df +1)}\,.
\label{eq:pMFT}
\ee
It is known that the asymptotic OPE density of heavy operators must be equal to that of mean field theory on average \cite{Qiao:2017xif,Mukhametzhanov:2018zja}. It follows that the heavy contributions to the sum rule \eqref{eq:sumRule} generally go like the power law $\Delta^{1-2k}$ at large $\Delta$. Thus, the sum rule is convergent only if $k>1$, as anticipated.

It will be convenient to define the heavy density $\omega[\Delta]$ associated to the functional $\omega$ as the LHS of \eqref{eq:heavyOmega}
\be
\omega[\Delta] = \frac{p^{\text{MFT}}_{\Delta}\,\omega[G^{\Df}_{\Delta}]}{2\sin^2\!\left[\tfrac{\pi}{2}(\Delta-2\Df)\right]}=
p^{\text{MFT}}_{\Delta}\int\limits_{-\infty}^{0}dz f(z)\,\widehat{G}^{\Df}_{\Delta}(z)
\,.
\label{eq:heavyDensity}
\ee
The contribution of operators with $\Delta>\Dgap$ to the sum rule \eqref{eq:sumRule} is then
\be
\omega|_{\text{UV}} = \sum\limits_{\Delta_{\cO}>\Dgap} \frac{(c_{\phi\phi\cO})^2}{p^{\text{MFT}}_{\Delta_{\cO}}} 2\sin^2\!\left[\tfrac{\pi}{2}(\Delta_{\cO}-2\Df)\right]\omega[\Delta_{\cO}]\,.
\ee
The definition of $\omega[\Delta]$ for fermionic functionals is the same, with $\cos$ instead of $\sin$.

The preceding discussion can be conveniently extended to a systematic expansion around the Regge limit by introducing the \emph{Regge moments}. Following \cite{Caron-Huot:2021enk}, we define the $k$th Regge moment of $\cG(z)$ as the weighted integral of $\dDisc\,\cG(z)$
\be
\Pi_k[\cG] = \int\limits_{0}^{1}d\rho \rho^{k-2}\dDisc\,\cG(z(\rho))\,.
\label{eq:Pik}
\ee
Here $\rho$ is the standard u-channel rho-coordinate \cite{Hogervorst:2013sma}, which is related to $z$ as follows
\be
\rho(z) = \frac{1}{(\sqrt{-z}+\sqrt{1-z})^2}\,,\qquad z(\rho) = -\frac{(1-\rho)^2}{4\rho}\,.
\ee
Any dispersive functional $\omega$ of Regge spin $k$ has a formal expansion in the Regge moments
\be
\Pi_{k},\,\Pi_{k+2}\,,\Pi_{k+4}, \ldots\,.
\ee
To see this, note that the measure $f(z)dz$ of the dispersive functional \eqref{eq:dispersiveF} can be expanded in  the Regge moment measures $\rho^{k-2}d\rho$
\be
f(z)dz = \sum\limits_{n=0}^{\infty}b_{n}\,\rho^{k+2n-2}d\rho\,,
\ee
which is just the Taylor series around the Regge limit $z=-\infty$. Since $f(z)=f(1-z)$, only even moments appear. The expansion of $\omega$ into Regge moments encodes the large-$\Delta$ asymptotics of the heavy action $\omega[\Delta]$
\be
\omega[\Delta] =
\sum_{n=0}^{N-1}b_n\,\Pi_{k+2n}[\Delta] 
+O(\Delta^{1-2k-4N})\quad\text{as}\quad\Delta\rightarrow\infty\,.
\label{eq:omegaExpansionPi}
\ee

It is important to stress that $\Pi_k$ is not a genuine functional, i.e. it does not give rise to an exact sum rule. This is because the kernel $f(z)$ corresponding to $\Pi_k$ does not satisfy the functional equation \eqref{eq:functionalEq}. However, we will see that $\Pi_k$ can be approximated using genuine functionals to arbitrarily high order around the Regge limit, effectively inverting the expansion \eqref{eq:omegaExpansionPi}.

\subsection{AdS$_2$ contact diagrams from Mellin space}\label{ssec:contacts}
The flat-space dispersive sum rules $\cC_k$ were chosen to be dual to the flat-space contact diagrams with amplitudes $\cM_{k}(s) = [s(s-4m^2)]^{\frac{k}{2}}$. One of our main goals is to construct analogous $\text{CFT}_1$ sum rules which are dual to the corresponding contact diagrams in $\text{AdS}_2$. These diagrams were analyzed previously in the nice paper \cite{Bianchi:2021piu}. Here we will review the constructions and OPE of these diagrams.

$\text{AdS}_2$ contact diagrams are precisely the four-point functions which are crossing-symmetric and admit an OPE of the form
\be
\cD(z) = \sum\limits_{n=0}^{\infty}\left[\delta a_n\,G_{\Delta_n}(z) +a_n \gamma_n \partial_{\Delta}G_{\Delta_n}(z)\right]\,,
\ee
with the anomalous dimensions $\gamma_n$ bounded by a power of $n$. Here $a_n$ are the squared OPE coefficients in mean field theory, i.e. $a_n = 2 p^{\text{MFT}}_{\Delta_n}$, where $p^{\text{MFT}}_{\Delta}$ is the OPE density of mean field theory, given in \eqref{eq:pMFT}. The formula $a_n = 2 p^{\text{MFT}}_{\Delta_n}$ applies also for fermions after the substitution $\Df\rightarrow\Dp$ and $\Delta_n=2\Dp+2n+1$.

We will define $\cD_k(z)$ to be the contact diagram that goes like spin $k$ in the Regge limit. For bosons, $\cD_k(z)$ corresponds to the bulk interaction with four $\Phi$s and $2k$ derivatives, i.e. $(\Phi\partial^{k}\Phi)^2$, where $k=0,2,4,\ldots$. For fermions, it corresponds to an interaction with four $\Psi$s and $2(k-1)$ derivatives, i.e. $(\Psi\partial^{k-1}\Psi)^2$, and we have $k=2,4,6,\ldots$. As $k$ ranges over these values, $\cD_k(z)$ produces a complete set of contact diagrams. Saying that $\cD_k(z)$ goes like spin $k$ leaves the ambiguity of adding linear combinations of contact diagrams with smaller $k$. We will fix this ambiguity by demanding
\ba
&\cD^{\rB}_k(z):\qquad \gamma_n = 0\text{ for all }n=0,\ldots,\tfrac{k}{2}-1\\
&\cD^{\rF}_k(z):\qquad \gamma_n = 0\text{ for all }n=0,\ldots,\tfrac{k}{2}-2\,.
\label{eq:Dgammas}
\ea
These conditions correspond to the fact that $\cM_{k}(s) = [s(s-4m^2)]^{\frac{k}{2}}$ has a zero of order $k/2$ at the threshold $s=4m^2$. Finally, we will fix the overall normalization of $\cD_k$ by demanding it goes to $\cM_{k}(s)$ in the bulk-point limit $n\rightarrow\infty$. At large $n$, the energy squared of the intermediate bulk state goes like $R^2 s\sim 4n^2$. At the same time, the scattering phase shift is related to the anomalous dimensions through the formula
\be
S(s) = 1 + \frac{i \cM(s)}{2\sqrt{s(s-4m^2)}} \sim e^{-i\pi\gamma_n}\,.
\ee
It follows that the anomalous dimension dual to the interaction $g_k\cM_k(s)$ go like
\be
\gamma_n\sim -g_kR^{2-2k}\frac{2^{2k-3}}{\pi}n^{2k-2}\quad\text{as}\quad n\rightarrow \infty\,.
\label{eq:gammaNormalizationFlat}
\ee
Thus, we will normalize $\cD_k$ to satisfy $\gamma_n\sim - (2^{2k-3}/\pi)n^{2k-2}$ as $n\rightarrow\infty$. Note that in this argument, we are not taking the flat space limit $R\rightarrow\infty$ but rather the bulk-point limit $n\rightarrow\infty$ at fixed AdS radius.

Contact diagrams $\cD^{\rB}_k(z)$,  $\cD^{\rF}_k(z)$ satisfying these properties can be conveniently represented as Mellin integrals. The general Mellin representation of bosonic AdS$_{2}$ contact diagrams takes the form
\ba
\cD^{\rB}(z) = \iint\!\!\frac{d\mS\,d\mT}{(4\pi i)^2}\,u^{\tfrac{\mS}{2}-\Df}v^{\tfrac{\mT}{2}-\Df}
\Gamma\!\left(\Df-\tfrac{\mS}{2}\right)^2\Gamma\!\left(\Df-\tfrac{\mT}{2}\right)^2\Gamma\!\left(\Df-\tfrac{\mU}{2}\right)^2 M^{\rB}(\mS,\mT)\,,
\label{eq:mellinB}
\ea
where we restrict to 1D CFT kinematics $u=z^2$, $v=(1-z)^2$. The Mellin variables satisfy $\mS+\mT+\mU = 4\Df$. The Mellin amplitude $M^{\rB}(\mS,\mT)$ is a polynomial satisfying
\be
M^{\rB}(\mS,\mT) = M^{\rB}(\mT,\mS) = M^{\rB}(\mS,\mU)\,.
\ee
The first equality ensures that $\cD^{\rB}(z)$ is crossing symmetric, and the second one ensures it only contains double-trace dimension $\Delta^{\rB}_n$ in its OPE (otherwise, it would also contain conformal blocks with dimensions $2\Df+n$ with $n$ odd).

There is no problem with using the two-variable Mellin representation \eqref{eq:mellinB} to describe 1D correlators, which only depend on a single cross-ratio. The absence of a unique inversion $\cD(z)\mapsto M(\mS,\mT)$ is of no concern here since we only use \eqref{eq:mellinB} as a convenient representation of $\cD(z)$.

The Mellin amplitude which gives rise to the contact diagram $\cD^{\rB}_k(z)$ is
\be
M^{\rB}_{k}(\mS,\mT) = c^{\rB}_k \left(\Df-\tfrac{\mS}{2}\right)_{\frac{k}{2}} \left(\Df-\tfrac{\mT}{2}\right)_{\frac{k}{2}} \left(\Df-\tfrac{\mU}{2}\right)_{\frac{k}{2}}\,.
\label{eq:MkBoson}
\ee
The Pochhammer symbols ensure the vanishing of the first $k/2$ anomalous dimensions as required by \eqref{eq:Dgammas}. $c^{\rB}_k$ is a normalization ensuring the correct flat-space limit
\be
c^{\rB}_k = (-1)^{\frac{k}{2}} \frac{2^{4 \Df+3 k-5} \Gamma (2 \Df+k -1) \Gamma \left(2 \Df+\frac{3 k}{2}-\frac{1}{2}\right)}{\pi ^{3/2} \Gamma (2 \Df )^2 \Gamma (2 \Df+2 k -1)}\,.
\ee
In AdS$_2$, the amplitude $M^{\rB}_{k}(\mS,\mT)$ gives rise to a contact diagram with $2k$ derivatives. This is in contrast with AdS$_{D>2}$, where it would have $3k$ derivatives. The difference is explained by noting that the Mandelstam variable $u$ vanishes in 2D flat space and therefore the last factor in \eqref{eq:MkBoson} does not contribute to the bulk mass dimension in $D=2$.

Fermionic contact diagrams can be treated almost identically. Their general Mellin representation reads
\ba
\cD^{\rF}(z) = \iint\!\!\frac{d\mS\,d\mT}{(4\pi i)^2}\,u^{\tfrac{\mS}{2}-\Dp}v^{\tfrac{\mT}{2}-\Dp}
\Gamma\!\left(\Dp+\tfrac{1-\mS}{2}\right)^2\Gamma\!\left(\Dp+\tfrac{1-\mT}{2}\right)^2\Gamma\!\left(\Dp+\tfrac{1-\mU}{2}\right)^2 M^{\rF}(\mS,\mT)\,,
\label{eq:mellinF}
\ea
where $u=z^2$, $v=(1-z)^2$ and $M^{\rF}(\mS,\mT)=M^{\rF}(\mT,\mS)=M^{\rF}(\mS,\mU)$. The Mellin amplitude giving rise to $\cD^{\rF}_k(z)$ reads
\be
M^{\rF}_{k}(\mS,\mT) = c^{\rF}_k \left(\Dp+\tfrac{1-\mS}{2}\right)_{\frac{k-2}{2}}\left(\Dp+\tfrac{1-\mT}{2}\right)_{\frac{k-2}{2}}\left(\Dp+\tfrac{1-\mU}{2}\right)_{\frac{k-2}{2}}\,,
\label{eq:MkBoson}
\ee
with normalization
\be
c^{\rF}_k = 
(-1)^{\frac{k}{2}+1}\frac{2^{4 \Dp +3 k-5} \Gamma (2 \Dp+k-1) \Gamma \left(2 \Dp+\frac{3 k}{2} -\frac{1}{2}\right)}{\pi ^{3/2} (2 \Dp +k-1)\Gamma (2 \Dp )^2 \Gamma (2 \Dp+2 k -2)}\,.
\ee
In Appendix \ref{app:gammas}, we give closed formulas for the anomalous dimension $\gamma_n$ of the contact diagrams $\cD^{\rB}_k(z)$ and $\cD^{\rF}_k(z)$ for general $k,n$. The formulas are in agreement with results obtained in \cite{Bianchi:2021piu}.

\section{Disperive sum rules and bounds in AdS$_2$}\label{sec:Dispersion}
\subsection{The $C_k$ dispersive functionals}
Let us now explain how to construct the dispersive functionals $C_k$, which will play the same role for AdS$_{2}$ scattering as the S-matrix sum rules $\cC_{k}$ play in flat space. Thus $C_k$ should have Regge spin $k$ and satisfy
\be
C^{\rB}_k\langle\cD^{\rB}_{k'}\rangle = -\delta_{kk'}\,,\qquad
C^{\rF}_k\langle\cD^{\rF}_{k'}\rangle= -\delta_{kk'}\,.
\label{eq:Cduality}
\ee
Here the notation $\omega\langle\cG\rangle$ means that we expand $\cG$ using the OPE and swap the sum over conformal blocks with the action of $\omega$. Thus in general we have $\omega\langle\cG\rangle\neq0$ even if $\cG$ is crossing symmetric, since $\omega$ may not commute with the OPE of $\cG$. Since $C_k$ has Regge spin $k$ and $\cD_{k'}$ has Regge spin $k'$, it follows that $C_k\langle\cD_{k'}\rangle = 0$ automatically for $k>k'$, but not for $k\leq k'$.

To construct $C_k$ satisfying the above properties, we can proceed as follows. In \cite{Mazac:2018ycv}, it was shown that there is a basis for the space of functionals for the 1D crossing equation \eqref{eq:bootstrapEq}, consisting of dispersive functionals $\alpha_n$, $\beta_n$. These functionals are dual to the double trace conformal blocks and their derivatives. In the bosonic case, this means
\ba
&\alpha^{\rB}_m[G^{\Df}_{\Delta^{\rB}_n}] = \delta_{mn} \qquad \beta^{\rB}_m[G^{\Df}_{\Delta^{\rB}_n}] = 0\\
&\alpha^{\rB}_m[\partial_{\Delta}G^{\Df}_{\Delta^{\rB}_n}] =0 \qquad \beta^{\rB}_m[\partial_{\Delta}G^{\Df}_{\Delta^{\rB}_n}] = \delta_{mn}\,,
\ea
where we use the notation $\partial_{\Delta}G^{\Df}_{\Delta^{\rB}_n} = (\partial_{\Delta}G^{\Df}_{\Delta})|_{\Delta=\Delta^{\rB}_n}$. Similarly, in the fermionic case, we have
\ba
&\alpha^{\rF}_m[G^{\Dp}_{\Delta^{\rF}_n}] = \delta_{mn} \qquad \beta^{\rF}_m[G^{\Dp}_{\Delta^{\rF}_n}] = 0\\
&\alpha^{\rF}_m[\partial_{\Delta}G^{\Dp}_{\Delta^{\rF}_n}] =0 \qquad \beta^{\rF}_m[\partial_{\Delta}G^{\Dp}_{\Delta^{\rF}_n}] = \delta_{mn}\,.
\ea
The functionals $\alpha^{\rB}_n$ and $\beta^{\rB}_n$ have Regge spin $k=0$ while $\alpha^{\rF}_n$ and $\beta^{\rF}_n$ have Regge spin $k=2$. We will construct $C^{\rB}_{k}$ and $C^{\rF}_{k}$ as finite linear combinations of the $\beta_n$ functionals
\be
C^{\rB}_k = \sum\limits_{n=0}^{\frac{k}{2}}x^{\rB}_{k,n}\beta^{\rB}_{n}\,,\qquad
C^{\rF}_k = \sum\limits_{n=0}^{\frac{k-2}{2}}x^{\rF}_{k,n}\beta^{\rF}_{n}\,.
\ee
The expansion coefficients $x^{\rB}_{k,n}$, $x^{\rF}_{k,n}$ are uniquely fixed by the conditions \eqref{eq:Cduality}. For example, the $k=k'$ conditions imply
\be
x^{\rB}_{k,k/2} = -\frac{1}{a^{\rB}_{k/2}\gamma^{\rB}_{k/2}}\,,\qquad
x^{\rF}_{k,(k-2)/2} = -\frac{1}{a^{\rF}_{(k-2)/2}\gamma^{\rF}_{(k-2)/2}}\,,
\ee
where $\gamma^{\rB}_{k/2}$ is the first nonvanishing anomalous dimension of $\cD^{\rB}_{k}$, and $\gamma^{\rF}_{(k-2)/2}$ the first nonvanishing anomalous dimension of $\cD^{\rF}_{k}$. The remaining $x_{k,n}$ can be fixed by demanding \eqref{eq:Cduality} for $k'<k$, or equivalently by demanding that $C_k$ has spin $k$ in the Regge limit, i.e. $f(z) = O(z^{-k})$. In both cases, the number of constraints equals the number of unknowns, and there is a unique solution. Note that \eqref{eq:Cduality} for $k'>k$ is automatic thanks to our definition of the contact diagrams, see \eqref{eq:Dgammas}. $C_{k}^{\rB}$ exists for $k=0,2,4,\ldots$ but only gives a nonperturbatively valid functional for $k>0$. On the other hand $C_{k}^{\rF}$ only exists for $k=2,4,6,\ldots$.

We will present the explicit construction of $C^{\rB}_k$ and $C^{\rF}_k$ using the $f(z)$ kernel appearing in \eqref{eq:omegaGeneral}. Let $f^{\rB}_{k}(z)$ and $f^{\rF}_{k}(z)$ be the kernels defining $C^{\rB}_k$ and $C^{\rF}_k$. We found
\ba
f^{\rB}_{k}(z) &= \frac{2 \Gamma (2 \Df )^2 \Gamma(\Df+\frac{k-1}{2}) \Gamma(\Df+k +\frac{1}{2})}{\pi  \Gamma (2 \Df+k -1)^2}\tfrac{1-2z}{[z(z-1)]^{\frac{k+1}{2}}}\times\\
&\times\left\{
{}_3\widetilde{F}_2\left(-\tfrac{k+1}{2},\tfrac{k+3}{2},2 \Df +\tfrac{3k-3}{2};\Df +\tfrac{k-1}{2},\Df +k+\tfrac{1}{2};\tfrac{1}{4 z (1-z)}\right)
+\right.\\
&\left.\quad+
\tfrac{3 (k+1) (k+3)}{16 z (z-1)}
{}_3\widetilde{F}_2\left(-\tfrac{k-1}{2},\tfrac{k+5}{2},2 \Df +\tfrac{3k-1}{2};\Df +\tfrac{k+1}{2},\Df +k+\tfrac{3}{2};\tfrac{1}{4 z (1-z)}\right)\right\}
\label{eq:fCkB}
\ea
and
\ba
f^{\rF}_{k}(z) &= \frac{2 \Gamma (2 \Dp )^2 \Gamma(\Dp+\frac{k}{2}) \Gamma(\Dp+k)}{\pi  \Gamma (2 \Dp+k -1)^2}\tfrac{1-2z}{[z(z-1)]^{\frac{k+1}{2}}}\times\\
&\times\left\{
{}_3\widetilde{F}_2\left(-\tfrac{k-1}{2},\tfrac{k+1}{2},2 \Dp +\tfrac{3k-3}{2};\Dp +\tfrac{k}{2},\Dp +k;\tfrac{1}{4 z (1-z)}\right)
+\right.\\
&\left.\quad+
\tfrac{3 (k-1) (k+1)}{16 z (z-1)}
{}_3\widetilde{F}_2\left(-\tfrac{k-3}{2},\tfrac{k+3}{2},2 \Dp +\tfrac{3k-1}{2};\Dp +\tfrac{k+2}{2},\Dp +k+1;\tfrac{1}{4 z (1-z)}\right)\right\}\,.
\label{eq:fCkF}
\ea
Here $_{3}\widetilde{F}_{2}$ is the regularized hypergeometric function. These formulas are valid for $z<0$ and are extended from there by analytic continuation. The analytic continuation satisfies $f(z)=f(1-z)$ as required. These results generalize the construction of $C_{2}^{\rF}\sim\beta^{\rF}_0$, given in \cite{Mazac:2018mdx}. In practice, the formulas were found in the same manner as the $f(z)$ kernel for $\beta_{0}^{\rF}$ was found in \cite{Mazac:2018mdx}. Namely, we first constructed $f(z)$ for many discrete values of $\Df\in\mathbb{N}$ and $\Dp\in\mathbb{N}+\tfrac{1}{2}$ by explitly solving the constraints. Next, we performed a Mellin transform of $f(z)$ and noticed that it admits a simple analytic continuation to general $\Df$, $\Dp$. Formulas \eqref{eq:fCkB}, \eqref{eq:fCkF} were then obtained by the inverse Mellin transform. See \cite{Mazac:2018mdx} for more details. One can check a posteriori that \eqref{eq:fCkB} and \eqref{eq:fCkF} satisfy all the constraints.

An important property of $f^{\rB}_{k}(z)$ and $f^{\rF}_{k}(z)$ is positivity
\be
f^{\rB}_{k}(z)> 0\quad\text{and}\quad f^{\rF}_{k}(z)>0\quad\text{for all}\quad z\in(-\infty,0)\,.
\ee
We expect this to be valid for all $\Df,\Dp>0$ and all $k\geq 2$. Although we have not found a general proof of positivity, we were able to verify it by numerically computing the kernels in many cases, and also by observing positivity of coefficients in their Taylor expansion around $z=\infty$. A direct consequence of positivity of $f(z)$ is positivity of action of $C_k$ on heavy conformal blocks. Indeed, it follows from \eqref{eq:dispersiveManifest} and $f(z)>0$ that
\ba
C^{\text{B}}_k[G_{\Delta}^{\Df}]&\geq 0 \quad\text{for all}\quad \Delta\geq 2\Df+k\\
C^{\text{F}}_k[G_{\Delta}^{\Dp}]&\geq 0 \quad\text{for all}\quad \Delta\geq 2\Df+k-1\,.
\ea
In summary, $C^{\text{B}}_k[G_{\Delta}^{\Df}]$ has simple zeros at $\Delta=2\Df+2n$ for all $0\leq n\leq k/2$, double zeros on all higher double traces, and is nonnegative for $\Delta \geq 2\Df+k$. Similarly, $C^{\text{F}}_k[G_{\Delta}^{\Dp}]$ has simple zeros at $\Delta=2\Dp+2n+1$ for all $0\leq n\leq (k-2)/2$, double zeros on all higher double traces, and is nonnegative for $\Delta \geq 2\Dp+k-1$.

In many applications, such as to prove the bounds at large $\Dgap$ later in this section, it is sufficient to use positivity of $f(z)$ at the leading order as $z\rightarrow -\infty$, which is manifestly true in \eqref{eq:fCkB} and \eqref{eq:fCkF}.

\subsection{AdS$_2$ arcs}
Having defined functionals $C_k$, we can proceed in parallel with the flat-space discussion of Section \ref{ssec:DispersionFlat}. The application of $C_k$ to the OPE of a crossing-symmetric correlator leads to the sum rule
\be
-C_k|_{\text{IR}} = C_k|_{\text{UV}}\,.
\label{eq:DispersiveCFT}
\ee
Here
\ba
C_k|_{\text{IR}} &= \sum\limits_{\Delta_{\cO}<\Dgap}(c_{\phi\phi\cO})^2 C_k[G^{\Df}_{\Delta_{\cO}}]\\
C_k|_{\text{UV}} &= \sum\limits_{\Delta_{\cO}\geq \Dgap}(c_{\phi\phi\cO})^2 C_k[G^{\Df}_{\Delta_{\cO}}]\,.
\ea
$\Dgap$ is a scaling dimension defining the split between what we call light and heavy operators. It is related to the scale $M$ of Section \ref{ssec:DispersionFlat} by $M^2R^2 = \Dgap(\Dgap-1)$. We demand $\Dgap>2\Df+k$ so that \eqref{eq:dispersiveManifest} converges for all $\Delta\geq \Dgap$, i.e. $C_k[G^{\Df}_{\Delta}]\geq 0$ for all $\Delta\geq\Dgap$ and $C_k[G^{\Df}_{\Delta}]$ has double zeros on all double-trace dimensions above $\Dgap$.

It is now natural to define the AdS$_2$ analogue of the arc variables $\cA_k(M)$ as minus the IR contribution to the $C_k$ sum rule
\be
A^{\text{B}}_k(\Dgap) = -\sum\limits_{\Delta_{\cO}<\Dgap}(c_{\phi\phi\cO})^2 C^{\text{B}}_k[G^{\Df}_{\Delta_{\cO}}]\,.
\label{eq:arcAdS}
\ee
The arcs $A^{\text{B}}_k(\Dgap)$ of any unitary solution to the 1d bootstrap equation \eqref{eq:bootstrapEq} satisfy various bounds as a result of the dispersive sum rule \eqref{eq:DispersiveCFT} and positivity of the UV contributions.

Before discussing these bounds in detail, let us explain how $A^{\text{B}}_k(\Dgap)$ can be computed in terms of an EFT in AdS$_2$. Let us assume that the physics up to scale $\Dgap$ is captured by an EFT containing a single bulk scalar field $\Phi$ dual to a boundary primary operator $\phi$. This means that the only primary states with dimensions $\Delta<\Dgap$ are the identity, and multi-trace operators built out of $\phi$ and derivatives. Furthermore, if the EFT is weakly-coupled up to scale $\Dgap$, the $\phi\times\phi$ OPE has a simple structure. At the zeroth order in the coupling, it contains only the identity and double-trace operators $[\phi\phi]_n$ with dimensions $2\Df+2n$. At the first order, the double traces acquire anomalous dimensions and anomalous OPEs. The most general tree-level four-point function is a combination of bulk contact diagrams $\cD^{\text{B}}_k$, discussed in Section \ref{ssec:contacts}
\be
\cG_{\text{B}}(z)|_{\text{tree}} = \sum\limits_{\substack{k=0\\k\text{ even}}}^{\infty} g_k \cD^{\text{B}}_k(z)\,.
\ee
It follows directly from the definition of the $C_k$ functionals \eqref{eq:Cduality} that at tree-level
\be
A^{\text{B}}_k(\Dgap) = g_k\,,
\ee
provided $\Dgap>2\Df+k$. In other words, the arc extracts the bulk four-point coupling $g_k$. At one-loop, the arc \eqref{eq:arcAdS} receives contributions linear in one-loop anomalous dimensions, as well as quadratic in the tree-level anomalous dimensions. The latter come from expanding $C^{\text{B}}_k[G^{\Df}_{\Delta}]$ to the second order around the double zeros at $\Delta = 2\Df+2n$. In flat space, these correspond to the imaginary part of $\cM(s)$ for $4m^2<s<M^2$, which indeed appears at one loop. Even at finite coupling, \eqref{eq:arcAdS} is a well-defined expression for the AdS arc in terms of quantities measurable by an observer who has access to bulk physics up to energy $\Dgap/R$.

Let us conclude this subsection by giving a formula for the action of a general dispersive functional $\omega$ on a general tree-level correlator $\cG(z)$. Thus, let $\omega$ be given by \eqref{eq:omegaGeneral} and let $\cG(z)$ be a crossing-symmetric combination of contact and exchange Witten diagrams. This means that $\omega\langle\cG\rangle$ only receives contributions from the finitely-many double-traces where $\omega$ does not have a double zero, as well as from the finitely-many single-trace operators coming from exchange diagrams. By deforming the contour in the $z$-plane from the left-hand branch cut to the right-hand branch cut, we get
\be
\omega\langle\cG\rangle = - \omega\langle\cG\rangle \pm 2\pi i [f(z)\cG(z)]_{z^{-1}}\,,
\ee
where the second term on the RHS is the contribution from the residue at $z=i\infty$ and the upper and lower sign corresponds to bosons and fermions. Notation $[f(z)\cG(z)]_{z^{-1}}$ means the coefficient of $z^{-1}$ as $z\rightarrow \infty$ in the upper half-plane. Thus,
\be
\omega\langle\cG\rangle = \pm i\pi [f(z)\cG(z)]_{z^{-1}}\,.
\label{eq:lightResidue}
\ee
This formula is the AdS analogue of evaluating the tree-level contributions to the flat-space arc \eqref{eq:CIR} as a residue at $s=\infty$. ln particular, we must have
\be
i\pi [f^{\text{B}}_k(z)\cD^{\text{B}}_{k'}(z)]_{z^{-1}} = -\delta_{kk'}\,,\qquad
i\pi [f^{\text{F}}_k(z)\cD^{\text{F}}_{k'}(z)]_{z^{-1}} = \delta_{kk'}\,,
\ee
which is a nontrivial statement about the $z\rightarrow i\infty$ expansion of the contact diagrams. It is also immediate from \eqref{eq:lightResidue} that tree-level couplings of $\Phi$ to other light matter fields do not contribute to the arcs. This is because exchange Witten diagrams in AdS$_{2}$ have Regge spin 0, and thus the residue \eqref{eq:lightResidue} vanishes.

\subsection{Action on heavy blocks}
Having discussed the light contributions to dispersive sum rules, let us turn to the contributions of heavy operators. We can use the explicit expressions \eqref{eq:fCkB}, \eqref{eq:fCkF} to find the expansion of $C^{\rB}_k$, $C^{\rF}_{k}$ into Regge moments to an arbitrary order. At the leading order, we have
\ba
C^{\rB}_{k} &=
\frac{4^k \Gamma (2 \Df )^2}{\pi\Gamma (2 \Df+k-1)^2}\Pi^{\rB}_{k} + O(\Pi^{\rB}_{k+2})\\
C^{\rF}_{k} &=
\frac{4^k \Gamma (2 \Dp )^2}{\pi  \Gamma (2 \Dp+k-1)^2}\Pi^{\rF}_{k} + O(\Pi^{\rF}_{k+2})\,.
\label{eq:CkRegge}
\ea
The notation $O(\Pi_{k+2})$ refers to suppression in the limit $\Delta\rightarrow\infty$, see \eqref{eq:omegaExpansionPi}. We see that arbitrary Regge moments with $k=2,4,\ldots$ can be approximated by linear combinations of functionals $C_k$ up to corrections which can be made arbitrarily small as $\Delta\rightarrow\infty$ by inverting the expansion \eqref{eq:CkRegge}. However, there is nothing particularly special about our definition Regge moments \eqref{eq:Pik}. We may have also defined the Regge moments to be precisely $C^{\rB}_{k}$, $C^{\rF}_{k}$, which have the advantage of being physical functionals.

It is instructive to evaluate the action of $C^{\rB}_{k}$, $C^{\rF}_{k}$ on heavy conformal blocks
\ba
C^{\rB}_{k}[\Delta] &= \frac{8}{\pi [\Delta(\Delta-1)]^{k-\frac{1}{2}}}\left[1+
\frac{48 \Df(\Df -1)  k-4 k^3+4 k+3}{24\Delta(\Delta-1)}
+O(\Delta^{-4})\right]\\[7pt]
C^{\rF}_{k}[\Delta] &= \frac{8}{\pi [\Delta(\Delta-1)]^{k-\frac{1}{2}}}\left[1+
\frac{48 \Dp(\Dp -1)  k-4 k^3+4 k+3}{24\Delta(\Delta-1)}
+O(\Delta^{-4})\right]\,.
\label{eq:CkHeavy}
\ea
Here $\omega[\Delta]$ is the heavy density of $\omega$, defined in \eqref{eq:heavyDensity}. One can check that the expansion proceeds in inverse powers of the Casimir $\Delta(\Delta-1)$. While $C^{\rB}_{k}[\Delta]$ and $C^{\rF}_{k}[\Delta]$ agree with each other to the order shown here, they differ already at the next order.

At the leading order as $\Delta\rightarrow\infty$, $C^{\rB}_{k}$ and $C^{\rF}_{k}$ exactly agree with the flat-space sum rules $\cC_{k}$. To see this, recall that the contribution of heavy states to $\cC_{k}$ is
\be
\cC_k|_{\text{UV}} = \int\limits_{M^2}^{\infty}\frac{ds}{\pi}\frac{2(s-2m^2)}{[s(s-4m^2)]^{\frac{k}{2}+1}}\mathrm{Im}[\cM(s)]\,.
\ee
If $M\gg m$, we can approximate this by
\be
\cC_k|_{\text{UV}} \sim \int\limits_{M^2}^{\infty}\frac{ds}{\pi}\frac{4}{s^{k}}\mathrm{Re}[1-S(s)]\,,
\label{eq:CFlatHeavy}
\ee
where we used \eqref{eq:SfromM}. On the other hand, from \eqref{eq:CkHeavy} we get for $\Dgap\gg 1$
\ba
C^{\rB}_{k}|_{\text{UV}} &=
\sum\limits_{\Delta_{\cO}>\Dgap} \frac{(c_{\phi\phi\cO})^2}{p^{\text{MFT}}_{\Delta_{\cO}}} 2\sin^2\!\left[\tfrac{\pi}{2}(\Delta_{\cO}-2\Df)\right]C^{\rB}_{k}[\Delta_{\cO}]\sim\\
&\sim
\sum\limits_{\Delta_{\cO}>\Dgap} \frac{(c_{\phi\phi\cO})^2}{p^{\text{MFT}}_{\Delta_{\cO}}} 2\sin^2\!\left[\tfrac{\pi}{2}(\Delta_{\cO}-2\Df)\right]
\frac{8}{\pi \Delta_{\cO}^{2k-1}}\,.
\label{eq:CHeavyExpansions}
\ea
This precisely agrees with \eqref{eq:CFlatHeavy} since $s\sim \Delta^2$ and thus $ds\sim 2\Delta\,d\Delta$. Furthermore, the S-matrix $S(s)$ corresponds to the local average of $e^{-i\pi(\Delta_{\cO}-2\Df)}$ weighted by the normalized OPE density $(c_{\phi\phi\cO})^2/p^{\text{MFT}}_{\Delta_{\cO}}$, i.e. for sufficiently large $\epsilon$
\ba
S(s) &\sim \frac{1}{2\epsilon}\sum\limits_{\sqrt{s}-\epsilon<\Delta_{\cO}<\sqrt{s}+\epsilon} \frac{(c_{\phi\phi\cO})^2}{p^{\text{MFT}}_{\Delta_{\cO}}} e^{-i\pi(\Delta_{\cO}-2\Df)}\\
\Rightarrow\quad\mathrm{Re}[1-S(s)] &\sim \frac{1}{2\epsilon}\sum\limits_{\sqrt{s}-\epsilon<\Delta_{\cO}<\sqrt{s}+\epsilon} \frac{(c_{\phi\phi\cO})^2}{p^{\text{MFT}}_{\Delta_{\cO}}} 2\sin^2\!\left[\tfrac{\pi}{2}(\Delta_{\cO}-2\Df)\right]\,.
\ea
It should not be a surprise that $C_{k}$ agrees with $\cC_{k}$ in the bulk-point limit since we have normalized the action of $C_{k}$ on AdS contact diagrams to agree with the action of $\cC_{k}$ on flat-space contact diagrams \eqref{eq:Cduality}, and our AdS contact diagrams have the correct flat-space normalization \eqref{eq:gammaNormalizationFlat}.

\subsection{Bounds on EFT couplings}
We are ready to derive bounds satisfied by the AdS arcs $A_k(\Dgap)$ as a consequence of the dispersive sum rules \eqref{eq:DispersiveCFT} and UV unitarity. For notational simplicity, we will assume that the tree-level approximation to the EFT is valid, and thus $A_k(\Dgap)=g_k$. All the bounds quoted remain valid at finite coupling after the replacement $g_k\rightarrow A_k(\Dgap)$.

It is an immediate consequence of positivity $C_k[G^{\Df}_{\Delta}]\geq 0$ for $\Delta\geq \Dgap$ that
\be
g_k \geq 0\,.
\ee
This bound is only saturated if all primary operators with $\Delta>\Dgap$ sit exactly at double-trace locations. Next, assuming $\Dgap$ is sufficiently large, so that the leading-order behaviour \eqref{eq:CkHeavy} can be trusted, we have
\be
0\leq\frac{g_\ell}{g_{k}}\leq \frac{C_\ell[\Dgap]}{C_{k}[\Dgap]}
\quad\text{for}\quad
2\leq k\leq \ell\,.
\label{eq:AdSUpper}
\ee
This bound is saturated by a single state at $\Delta=\Dgap$ and follows from
\be
\frac{C_{\ell}[\Delta]}{C_{\ell}[\Dgap]}\leq
\frac{C_{k}[\Delta]}{C_{k}[\Dgap]}
\quad\text{for}\quad
k\leq \ell\quad\text{and}\quad \Delta\geq\Dgap\,.
\ee
Let us expand the upper bound \eqref{eq:AdSUpper} at large $\Dgap$ using \eqref{eq:CkHeavy}
\be
0\leq\frac{g_\ell}{g_{k}}\leq \frac{1}{M^{2(\ell-k)}}\left[1+
\frac{(\ell-k)(12 \Df (\Df-1) -k^2-k \ell-\ell^2+1)}{6 M^2}
+O(M^{-4})
\right]\,.
\label{eq:AdSUpper2}
\ee
Here $M^2 = \Dgap(\Dgap-1)$, i.e. we set $R_{\text{AdS}}=1$. To this order in $1/M^2$, the fermionic result is the same after $\Df\rightarrow\Dp$. We see that the bound agrees with the flat-space bound \eqref{eq:upperFlat} at the leading order in the bulk-point limit. The subleading order in \eqref{eq:AdSUpper2} constitutes a universal correction coming from the finite size of AdS.

We expect that \eqref{eq:AdSUpper} exactly reproduces \eqref{eq:upperFlat} at finite $M$ and $m$ by taking the flat-space limit $\Df,\Dgap\rightarrow\infty$, with $\Df/\Dgap = m/M$. Indeed, the piece containing $\Df(\Df-1)=m^2$ in \eqref{eq:AdSUpper2} agrees with the leading correction in $m^2/M^2$ of \eqref{eq:upperFlat}.

We can also derive the AdS version of the flat-space bound \eqref{eq:lowerFlat}. To do that, note that the dispersive sum rules can be stated as
\be
\frac{g_k}{g_2} = \int\limits_{\Dgap}^{\infty}\!\!\!\! d\Delta\, w(\Delta) \widehat{C}_k[\Delta]\,,\quad\text{where}\quad w(\Delta)\geq 0 \quad\text{and}\quad
\int\limits_{\Dgap}^{\infty}\!\!\!\! d\Delta\, w(\Delta) = 1\,,
\ee
where $\widehat{C}_k[\Delta] = C_k[\Delta]/C_2[\Delta]$. Suppose $2<k\leq \ell$ and $\Dgap\gg 1$, so that \eqref{eq:AdSUpper2} implies $g_k/g_2\ll1$ and $g_{\ell}/g_2\ll 1$. Consider the function $\widehat{C}_{\ell}\circ\widehat{C}^{-1}_k$, where $\widehat{C}^{-1}_k$ is the inverse of $\widehat{C}_k[\Delta]$. This function is approximately equal to
\be
\widehat{C}_{\ell}\circ\widehat{C}^{-1}_k(x) = x^{\frac{\ell-2}{k-2}}\left[1-
\frac{ (\ell-2) (\ell-k) (k+\ell+2)}{6}x^{\frac{1}{k-2}}+O(x^{\frac{2}{k-2}})\right]\quad\text{as}\quad x\rightarrow 0\,.
\ee
Therefore, for $\Dgap\gg1$, $\widehat{C}_{\ell}\circ\widehat{C}^{-1}_k$ is convex on the range of $\widehat{C}_{k}$ since $x^{\frac{\ell-2}{k-2}}$ is convex. Jensen's inequality, reviewed around \eqref{eq:jensen}, then implies the bound
\be
\left(\frac{g_{\ell}}{g_{2}}\right)^{k-2}\geq
\left(\frac{g_{k}}{g_{2}}\right)^{\ell-2}\left[1-\frac{(k-2) (\ell-2) (\ell-k) (k+l+2)}{6}\left(\frac{g_{k}}{g_{2}}\right)^{\frac{1}{k-2}}+O(M^{-4})\right]\,.
\label{eq:boundAdSLower}
\ee
We can see that the leading finite AdS radius correction slightly weakens the flat-space bound \eqref{eq:lowerFlat}. The bound is saturated by a single state with varying dimension $\Delta>\Dgap$.

\section{Conclusions and future directions}\label{sec:conclusions}
In this note, we studied 1D CFTs dual to 2D QFTs in AdS. We developed dispersive methods that allowed us to extract and put bounds on higher-derivative contact interactions. This was done by constructing a set of CFT functionals $C_k$, labelled by their Regge spin $k$. When we apply functional $C_k$ to a tree-level correlator in AdS$_2$, it extracts the bulk coupling of $(\Phi\partial^k\Phi)^2$ in the bosonic case and $(\Psi\partial^{k-1}\Psi)^2$ in the fermionic case.

The action of the dispersive functionals $C_k$ on a general 1D CFT correlator is given by
\begin{equation}
 C_k[\cG] = \int\limits_{-\infty}^{0}\!\! dz f_k(z)\, \dDisc\,\cG(z)\,,
\end{equation}
where the kernels $f_k(z)$ appear in \eqref{eq:fCkB} and \eqref{eq:fCkF}. This formula demonstrates that the functionals are dispersive, i.e.\ that they compute moments of the double commutator.

If we assume that the bulk theory is described by an EFT up to scale $\Dgap$, the application of $C_k$ to the conformal bootstrap equation relates the bulk higher-derivative couplings to positive averages of the heavy OPE density for $\Delta>\Dgap$, weighted by $\sin^2[\frac{\pi}{2}(\Delta-2\Df)]\Delta^{-2k}$ at large $\Delta$, see \eqref{eq:CHeavyExpansions}. This leads to bounds on ratios of the higher-derivative couplings depending on $\Dgap$. As $\Dgap\rightarrow\infty$, these bounds reproduce the flat space bounds following from unitarity and causality of the S-matrix. We computed the first subleading correction to the bounds at large $\Dgap$, coming from the finite size of AdS, see \eqref{eq:AdSUpper2} and \eqref{eq:boundAdSLower}.

We conclude by several suggestions for further work. We have studied the bounds in the bulk-point limit $\Dgap\rightarrow \infty$ at fixed $\Df$. This corresponds to high-energy scattering in AdS of finite size. It would be interesting to consider also the flat-space limit $\Dgap,\Df\rightarrow\infty$ at fixed $\Df/\Dgap = m/M$. We expect that in this limit, the AdS bound \eqref{eq:AdSUpper} will exactly reproduce the flat-space bound \eqref{eq:upperFlat} and the leading correction away from the flat space limit should be computable.

It would also be interesting to come up with a more conceptual derivation of the dispersive kernels $f_k(z)$ given in \eqref{eq:fCkB} and \eqref{eq:fCkF}. In higher dimensions, dispersive sum rules arise by integrating the crossing equation along a pair of null rays. While there are no null rays in 1D CFTs, there are null rays in AdS$_2$. Could the relatively complicated formulas for $f_k(z)$ admit a simple representation as integrals in AdS$_2$? In a similar spirit, dispersion relations in higher-dimensional CFTs take a simple form in Mellin space \cite{Penedones:2019tng,Caron-Huot:2020adz,Gopakumar:2021dvg}. Can the Mellin space derivation be extended to 1D CFTs?

\section*{Acknowledgments}
We would like to thank Joan Elias Miró and Leonardo Rastelli for useful discussions. WK is supported by the NSF grant PHY-1915093. DM acknowledges funding provided by Edward and Kiyomi Baird as well as the grant DE-SC0009988 from the U.S. Department of Energy.

\appendix
\section{Anomalous dimensions in general contact diagrams}\label{app:gammas}
One can find the anomalous dimensions of the contact diagrams $(\Phi\partial^{k}\Phi)^2$ and $(\Psi\partial^{k-1}\Psi)^2$, starting from their Mellin representations \eqref{eq:mellinB}, \eqref{eq:mellinF} as follows. First, we close the $\mS$ contour to the right, picking terms proportional to $\log(z)$ from the double poles at $\mS = 2\Df + 2n$. For each residue, the remaining integral over $\mT$ can be evaluated in terms of a ${}_2F_1$ hypergeometric function. The final step is to reorgranize the sum over $n$ in terms of 1D conformal blocks to read off the anomalous dimensions. By computing $\gamma_{n}$ in this manner explicitly in many examples, we were able to guess the formula for $\gamma_n$ of a general contact diagram. The anomalous dimension of $\cD^{\rB}_k(z)$ are
\ba
\gamma^{\rB}_{n} = 
&-\frac{\Gamma \left(\Df+\frac{k}{2}\right) \Gamma (2 \Df +k-1) \Gamma \left(2 \Df+\frac{3 k}{2} -\frac{1}{2}\right)}{2^{2 \Df+1} \sqrt{\pi } \Gamma \left(\Df +k-\frac{1}{2}\right)}\times\\
&\times\frac{\Gamma \left(n+\frac{1}{2}\right)\Gamma (\Df+n ) \Gamma \left(\Df -\frac{k}{2}+n\right) \Gamma \left(2 \Df+\frac{k}{2}+n -\frac{1}{2}\right)}{\Gamma \left(n-\frac{k}{2}+1\right) \Gamma \left(\Df +n+\frac{1}{2}\right) \Gamma (2 \Df +n) \Gamma \left(\Df+\frac{k}{2}+n +\frac{1}{2}\right)}\times\\
&\times
{}_4\widetilde{F}_3\left(-k,-n,2 \Df +k-1,2 \Df +n-\tfrac{1}{2};\Df ,\Df -\tfrac{k}{2},2 \Df +\tfrac{k}{2}-\tfrac{1}{2};1\right)\,.
\ea
Here ${}_4\widetilde{F}_3$ is the regularized hypergeometric function. The anomalous dimension of $\cD^{\rF}_k(z)$ are
\ba
\gamma^{\rF}_{n} = 
&-\frac{\Gamma \left(\Dp+\frac{k-1}{2} \right) \Gamma (2 \Dp+k -1) \Gamma \left(2 \Dp+\frac{3 k}{2} -\frac{1}{2}\right)}{2^{2 \Dp +3} \sqrt{\pi } \Gamma (\Dp +k-1)}\times\\
&\times
\frac{\Gamma \left(n+\frac{3}{2}\right)\Gamma \left(\Dp+n +\frac{1}{2}\right) \Gamma \left(\Dp-\frac{k}{2}+n +\frac{1}{2}\right) \Gamma \left(2 \Dp+\frac{k}{2}+n -\frac{1}{2}\right)}{\Gamma \left(n-\frac{k}{2}+2\right) \Gamma (\Dp+n +1) \Gamma (2 \Dp +n) \Gamma \left(\Dp+\frac{k}{2}+n +1\right)}\times\\
&\times\left[
n (4 \Dp +2 n+1)(4 \Dp  k-2 \Dp +2 k^2-4 k+3) A +2 (2\Dp+ 1)B\right]\,.
\ea
where
\ba
A &= {}_4\widetilde{F}_3\left(1-k,1-n,2 \Dp +n+\tfrac{3}{2},2 \Dp +k-1;\Dp +\tfrac{3}{2},\Dp+\tfrac{3-k}{2},2 \Dp +\tfrac{k+1}{2};1\right)\\
B &= {}_4\widetilde{F}_3\left(-k,-n,2 \Dp +n+\tfrac{1}{2},2 \Dp +k-2;\Dp +\tfrac{1}{2},\Dp +\tfrac{1-k}{2},2\Dp +\tfrac{k-1}{2};1\right)\,.
\ea

\newpage
\bibliography{refs}

\providecommand{\href}[2]{#2}\begingroup\raggedright\begin{thebibliography}{10}

\bibitem{Poland:2018epd}
D.~Poland, S.~Rychkov and A.~Vichi, \emph{{The Conformal Bootstrap: Theory,
  Numerical Techniques, and Applications}},
  \href{https://doi.org/10.1103/RevModPhys.91.015002}{\emph{Rev. Mod. Phys.}
  {\bfseries 91} (2019) 015002}
  [\href{https://arxiv.org/abs/1805.04405}{{\ttfamily 1805.04405}}].

\bibitem{Hartman:2022zik}
T.~Hartman, D.~Mazac, D.~Simmons-Duffin and A.~Zhiboedov, \emph{{Snowmass White
  Paper: The Analytic Conformal Bootstrap}},  in \emph{{2022 Snowmass Summer
  Study}}, 2, 2022 [\href{https://arxiv.org/abs/2202.11012}{{\ttfamily
  2202.11012}}].

\bibitem{Poland:2022qrs}
D.~Poland and D.~Simmons-Duffin, \emph{{Snowmass White Paper: The Numerical
  Conformal Bootstrap}},  in \emph{{2022 Snowmass Summer Study}}, 3, 2022
  [\href{https://arxiv.org/abs/2203.08117}{{\ttfamily 2203.08117}}].

\bibitem{Gopakumar:2022kof}
R.~Gopakumar, E.~Perlmutter, S.S.~Pufu and X.~Yin, \emph{{Snowmass White Paper:
  Bootstrapping String Theory}},
  \href{https://arxiv.org/abs/2202.07163}{{\ttfamily 2202.07163}}.

\bibitem{Kruczenski:2022lot}
M.~Kruczenski, J.~Penedones and B.C.~van Rees, \emph{{Snowmass White Paper:
  S-matrix Bootstrap}},  \href{https://arxiv.org/abs/2203.02421}{{\ttfamily
  2203.02421}}.

\bibitem{Weinberg:1978kz}
S.~Weinberg, \emph{{Phenomenological Lagrangians}},
  \href{https://doi.org/10.1016/0378-4371(79)90223-1}{\emph{Physica A}
  {\bfseries 96} (1979) 327}.

\bibitem{Weinberg:1995mt}
S.~Weinberg, \emph{{The Quantum theory of fields. Vol. 1: Foundations}},
  Cambridge University Press (2005).

\bibitem{Weinberg:1996kr}
S.~Weinberg, \emph{{The quantum theory of fields. Vol. 2: Modern
  applications}}, Cambridge University Press (2013).

\bibitem{Heemskerk:2009pn}
I.~Heemskerk, J.~Penedones, J.~Polchinski and J.~Sully, \emph{{Holography from
  Conformal Field Theory}},
  \href{https://doi.org/10.1088/1126-6708/2009/10/079}{\emph{JHEP} {\bfseries
  10} (2009) 079} [\href{https://arxiv.org/abs/0907.0151}{{\ttfamily
  0907.0151}}].

\bibitem{Fitzpatrick:2010zm}
A.L.~Fitzpatrick, E.~Katz, D.~Poland and D.~Simmons-Duffin, \emph{{Effective
  Conformal Theory and the Flat-Space Limit of AdS}},
  \href{https://doi.org/10.1007/JHEP07(2011)023}{\emph{JHEP} {\bfseries 07}
  (2011) 023} [\href{https://arxiv.org/abs/1007.2412}{{\ttfamily 1007.2412}}].

\bibitem{Fitzpatrick:2011dm}
A.L.~Fitzpatrick and J.~Kaplan, \emph{{Unitarity and the Holographic
  S-Matrix}}, \href{https://doi.org/10.1007/JHEP10(2012)032}{\emph{JHEP}
  {\bfseries 10} (2012) 032} [\href{https://arxiv.org/abs/1112.4845}{{\ttfamily
  1112.4845}}].

\bibitem{Pham:1985cr}
T.N.~Pham and T.N.~Truong, \emph{{Evaluation of the Derivative Quartic Terms of
  the Meson Chiral Lagrangian From Forward Dispersion Relation}},
  \href{https://doi.org/10.1103/PhysRevD.31.3027}{\emph{Phys. Rev. D}
  {\bfseries 31} (1985) 3027}.

\bibitem{Adams:2006sv}
A.~Adams, N.~Arkani-Hamed, S.~Dubovsky, A.~Nicolis and R.~Rattazzi,
  \emph{{Causality, analyticity and an IR obstruction to UV completion}},
  \href{https://doi.org/10.1088/1126-6708/2006/10/014}{\emph{JHEP} {\bfseries
  10} (2006) 014} [\href{https://arxiv.org/abs/hep-th/0602178}{{\ttfamily
  hep-th/0602178}}].

\bibitem{Camanho:2014apa}
X.O.~Camanho, J.D.~Edelstein, J.~Maldacena and A.~Zhiboedov, \emph{{Causality
  Constraints on Corrections to the Graviton Three-Point Coupling}},
  \href{https://doi.org/10.1007/JHEP02(2016)020}{\emph{JHEP} {\bfseries 02}
  (2016) 020} [\href{https://arxiv.org/abs/1407.5597}{{\ttfamily 1407.5597}}].

\bibitem{Hartman:2015lfa}
T.~Hartman, S.~Jain and S.~Kundu, \emph{{Causality Constraints in Conformal
  Field Theory}}, \href{https://doi.org/10.1007/JHEP05(2016)099}{\emph{JHEP}
  {\bfseries 05} (2016) 099}
  [\href{https://arxiv.org/abs/1509.00014}{{\ttfamily 1509.00014}}].

\bibitem{Baumann:2015nta}
D.~Baumann, D.~Green, H.~Lee and R.A.~Porto, \emph{{Signs of Analyticity in
  Single-Field Inflation}},
  \href{https://doi.org/10.1103/PhysRevD.93.023523}{\emph{Phys. Rev. D}
  {\bfseries 93} (2016) 023523}
  [\href{https://arxiv.org/abs/1502.07304}{{\ttfamily 1502.07304}}].

\bibitem{Bellazzini:2015cra}
B.~Bellazzini, C.~Cheung and G.N.~Remmen, \emph{{Quantum Gravity Constraints
  from Unitarity and Analyticity}},
  \href{https://doi.org/10.1103/PhysRevD.93.064076}{\emph{Phys. Rev. D}
  {\bfseries 93} (2016) 064076}
  [\href{https://arxiv.org/abs/1509.00851}{{\ttfamily 1509.00851}}].

\bibitem{Bonifacio:2016wcb}
J.~Bonifacio, K.~Hinterbichler and R.A.~Rosen, \emph{{Positivity constraints
  for pseudolinear massive spin-2 and vector Galileons}},
  \href{https://doi.org/10.1103/PhysRevD.94.104001}{\emph{Phys. Rev. D}
  {\bfseries 94} (2016) 104001}
  [\href{https://arxiv.org/abs/1607.06084}{{\ttfamily 1607.06084}}].

\bibitem{deRham:2017avq}
C.~de~Rham, S.~Melville, A.J.~Tolley and S.-Y.~Zhou, \emph{{Positivity bounds
  for scalar field theories}},
  \href{https://doi.org/10.1103/PhysRevD.96.081702}{\emph{Phys. Rev. D}
  {\bfseries 96} (2017) 081702}
  [\href{https://arxiv.org/abs/1702.06134}{{\ttfamily 1702.06134}}].

\bibitem{Caron-Huot:2022ugt}
S.~Caron-Huot, Y.-Z.~Li, J.~Parra-Martinez and D.~Simmons-Duffin,
  \emph{{Causality constraints on corrections to Einstein gravity}},
  \href{https://arxiv.org/abs/2201.06602}{{\ttfamily 2201.06602}}.

\bibitem{Bellazzini:2020cot}
B.~Bellazzini, J.~Elias~Mir\'o, R.~Rattazzi, M.~Riembau and F.~Riva,
  \emph{{Positive moments for scattering amplitudes}},
  \href{https://doi.org/10.1103/PhysRevD.104.036006}{\emph{Phys. Rev. D}
  {\bfseries 104} (2021) 036006}
  [\href{https://arxiv.org/abs/2011.00037}{{\ttfamily 2011.00037}}].

\bibitem{Arkani-Hamed:2020blm}
N.~Arkani-Hamed, T.-C.~Huang and Y.-T.~Huang, \emph{{The EFT-Hedron}},
  \href{https://doi.org/10.1007/JHEP05(2021)259}{\emph{JHEP} {\bfseries 05}
  (2021) 259} [\href{https://arxiv.org/abs/2012.15849}{{\ttfamily
  2012.15849}}].

\bibitem{Tolley:2020gtv}
A.J.~Tolley, Z.-Y.~Wang and S.-Y.~Zhou, \emph{{New positivity bounds from full
  crossing symmetry}},
  \href{https://doi.org/10.1007/JHEP05(2021)255}{\emph{JHEP} {\bfseries 05}
  (2021) 255} [\href{https://arxiv.org/abs/2011.02400}{{\ttfamily
  2011.02400}}].

\bibitem{Caron-Huot:2020cmc}
S.~Caron-Huot and V.~Van~Duong, \emph{{Extremal Effective Field Theories}},
  \href{https://doi.org/10.1007/JHEP05(2021)280}{\emph{JHEP} {\bfseries 05}
  (2021) 280} [\href{https://arxiv.org/abs/2011.02957}{{\ttfamily
  2011.02957}}].

\bibitem{Caron-Huot:2021rmr}
S.~Caron-Huot, D.~Mazac, L.~Rastelli and D.~Simmons-Duffin, \emph{{Sharp
  Boundaries for the Swampland}},
  \href{https://doi.org/10.1007/jhep07(2021)110}{\emph{JHEP} {\bfseries 07}
  (2021) 110} [\href{https://arxiv.org/abs/2102.08951}{{\ttfamily
  2102.08951}}].

\bibitem{Carmi:2019cub}
D.~Carmi and S.~Caron-Huot, \emph{{A Conformal Dispersion Relation:
  Correlations from Absorption}},
  \href{https://doi.org/10.1007/JHEP09(2020)009}{\emph{JHEP} {\bfseries 09}
  (2020) 009} [\href{https://arxiv.org/abs/1910.12123}{{\ttfamily
  1910.12123}}].

\bibitem{Mazac:2019shk}
D.~Mazac, L.~Rastelli and X.~Zhou, \emph{{A basis of analytic functionals for
  CFTs in general dimension}},
  \href{https://doi.org/10.1007/JHEP08(2021)140}{\emph{JHEP} {\bfseries 08}
  (2021) 140} [\href{https://arxiv.org/abs/1910.12855}{{\ttfamily
  1910.12855}}].

\bibitem{Kologlu:2019bco}
M.~Kologlu, P.~Kravchuk, D.~Simmons-Duffin and A.~Zhiboedov, \emph{{Shocks,
  Superconvergence, and a Stringy Equivalence Principle}},
  \href{https://doi.org/10.1007/JHEP11(2020)096}{\emph{JHEP} {\bfseries 11}
  (2020) 096} [\href{https://arxiv.org/abs/1904.05905}{{\ttfamily
  1904.05905}}].

\bibitem{Mazac:2016qev}
D.~Mazac, \emph{{Analytic bounds and emergence of AdS$_{2}$ physics from the
  conformal bootstrap}},
  \href{https://doi.org/10.1007/JHEP04(2017)146}{\emph{JHEP} {\bfseries 04}
  (2017) 146} [\href{https://arxiv.org/abs/1611.10060}{{\ttfamily
  1611.10060}}].

\bibitem{Mazac:2018mdx}
D.~Mazac and M.F.~Paulos, \emph{{The analytic functional bootstrap. Part I: 1D
  CFTs and 2D S-matrices}},
  \href{https://doi.org/10.1007/JHEP02(2019)162}{\emph{JHEP} {\bfseries 02}
  (2019) 162} [\href{https://arxiv.org/abs/1803.10233}{{\ttfamily
  1803.10233}}].

\bibitem{Mazac:2018ycv}
D.~Mazac and M.F.~Paulos, \emph{{The analytic functional bootstrap. Part II.
  Natural bases for the crossing equation}},
  \href{https://doi.org/10.1007/JHEP02(2019)163}{\emph{JHEP} {\bfseries 02}
  (2019) 163} [\href{https://arxiv.org/abs/1811.10646}{{\ttfamily
  1811.10646}}].

\bibitem{Paulos:2019gtx}
M.F.~Paulos, \emph{{Analytic functional bootstrap for CFTs in $d > 1$}},
  \href{https://doi.org/10.1007/JHEP04(2020)093}{\emph{JHEP} {\bfseries 04}
  (2020) 093} [\href{https://arxiv.org/abs/1910.08563}{{\ttfamily
  1910.08563}}].

\bibitem{Gopakumar:2016cpb}
R.~Gopakumar, A.~Kaviraj, K.~Sen and A.~Sinha, \emph{{A Mellin space approach
  to the conformal bootstrap}},
  \href{https://doi.org/10.1007/JHEP05(2017)027}{\emph{JHEP} {\bfseries 05}
  (2017) 027} [\href{https://arxiv.org/abs/1611.08407}{{\ttfamily
  1611.08407}}].

\bibitem{Penedones:2019tng}
J.~Penedones, J.A.~Silva and A.~Zhiboedov, \emph{{Nonperturbative Mellin
  Amplitudes: Existence, Properties, Applications}},
  \href{https://doi.org/10.1007/JHEP08(2020)031}{\emph{JHEP} {\bfseries 08}
  (2020) 031} [\href{https://arxiv.org/abs/1912.11100}{{\ttfamily
  1912.11100}}].

\bibitem{Carmi:2020ekr}
D.~Carmi, J.~Penedones, J.A.~Silva and A.~Zhiboedov, \emph{{Applications of
  dispersive sum rules: $\epsilon$-expansion and holography}},
  \href{https://doi.org/10.21468/SciPostPhys.10.6.145}{\emph{SciPost Phys.}
  {\bfseries 10} (2021) 145}
  [\href{https://arxiv.org/abs/2009.13506}{{\ttfamily 2009.13506}}].

\bibitem{Gopakumar:2021dvg}
R.~Gopakumar, A.~Sinha and A.~Zahed, \emph{{Crossing Symmetric Dispersion
  Relations for Mellin Amplitudes}},
  \href{https://doi.org/10.1103/PhysRevLett.126.211602}{\emph{Phys. Rev. Lett.}
  {\bfseries 126} (2021) 211602}
  [\href{https://arxiv.org/abs/2101.09017}{{\ttfamily 2101.09017}}].

\bibitem{Caron-Huot:2020adz}
S.~Caron-Huot, D.~Mazac, L.~Rastelli and D.~Simmons-Duffin, \emph{{Dispersive
  CFT Sum Rules}}, \href{https://doi.org/10.1007/JHEP05(2021)243}{\emph{JHEP}
  {\bfseries 05} (2021) 243}
  [\href{https://arxiv.org/abs/2008.04931}{{\ttfamily 2008.04931}}].

\bibitem{Caron-Huot:2021enk}
S.~Caron-Huot, D.~Mazac, L.~Rastelli and D.~Simmons-Duffin, \emph{{AdS bulk
  locality from sharp CFT bounds}},
  \href{https://doi.org/10.1007/JHEP11(2021)164}{\emph{JHEP} {\bfseries 11}
  (2021) 164} [\href{https://arxiv.org/abs/2106.10274}{{\ttfamily
  2106.10274}}].

\bibitem{Paulos:2016fap}
M.F.~Paulos, J.~Penedones, J.~Toledo, B.C.~van Rees and P.~Vieira, \emph{{The
  S-matrix bootstrap. Part I: QFT in AdS}},
  \href{https://doi.org/10.1007/JHEP11(2017)133}{\emph{JHEP} {\bfseries 11}
  (2017) 133} [\href{https://arxiv.org/abs/1607.06109}{{\ttfamily
  1607.06109}}].

\bibitem{Antunes:2021abs}
A.~Antunes, M.S.~Costa, J.~Penedones, A.~Salgarkar and B.C.~van Rees,
  \emph{{Towards bootstrapping RG flows: sine-Gordon in AdS}},
  \href{https://doi.org/10.1007/JHEP12(2021)094}{\emph{JHEP} {\bfseries 12}
  (2021) 094} [\href{https://arxiv.org/abs/2109.13261}{{\ttfamily
  2109.13261}}].

\bibitem{Liendo:2018ukf}
P.~Liendo, C.~Meneghelli and V.~Mitev, \emph{{Bootstrapping the half-BPS line
  defect}}, \href{https://doi.org/10.1007/JHEP10(2018)077}{\emph{JHEP}
  {\bfseries 10} (2018) 077}
  [\href{https://arxiv.org/abs/1806.01862}{{\ttfamily 1806.01862}}].

\bibitem{Mazac:2018qmi}
D.~Mazac, \emph{{A Crossing-Symmetric OPE Inversion Formula}},
  \href{https://doi.org/10.1007/JHEP06(2019)082}{\emph{JHEP} {\bfseries 06}
  (2019) 082} [\href{https://arxiv.org/abs/1812.02254}{{\ttfamily
  1812.02254}}].

\bibitem{Hartman:2019pcd}
T.~Hartman, D.~Mazac and L.~Rastelli, \emph{{Sphere Packing and Quantum
  Gravity}}, \href{https://doi.org/10.1007/JHEP12(2019)048}{\emph{JHEP}
  {\bfseries 12} (2019) 048}
  [\href{https://arxiv.org/abs/1905.01319}{{\ttfamily 1905.01319}}].

\bibitem{Paulos:2020zxx}
M.F.~Paulos, \emph{{Dispersion relations and exact bounds on CFT correlators}},
   \href{https://arxiv.org/abs/2012.10454}{{\ttfamily 2012.10454}}.

\bibitem{Ferrero:2021bsb}
P.~Ferrero and C.~Meneghelli, \emph{{Bootstrapping the half-BPS line defect CFT
  in N=4 supersymmetric Yang-Mills theory at strong coupling}},
  \href{https://doi.org/10.1103/PhysRevD.104.L081703}{\emph{Phys. Rev. D}
  {\bfseries 104} (2021) L081703}
  [\href{https://arxiv.org/abs/2103.10440}{{\ttfamily 2103.10440}}].

\bibitem{Cordova:2022pbl}
L.~C\'ordova, Y.~He and M.F.~Paulos, \emph{{From conformal correlators to
  analytic S-matrices: CFT$_1$/QFT$_2$}},
  \href{https://arxiv.org/abs/2203.10840}{{\ttfamily 2203.10840}}.

\bibitem{EliasMiro:2019kyf}
J.~Elias~Mir\'o, A.L.~Guerrieri, A.~Hebbar, J.a.~Penedones and P.~Vieira,
  \emph{{Flux Tube S-matrix Bootstrap}},
  \href{https://doi.org/10.1103/PhysRevLett.123.221602}{\emph{Phys. Rev. Lett.}
  {\bfseries 123} (2019) 221602}
  [\href{https://arxiv.org/abs/1906.08098}{{\ttfamily 1906.08098}}].

\bibitem{EliasMiro:2021nul}
J.~Elias~Mir\'o and A.~Guerrieri, \emph{{Dual EFT bootstrap: QCD flux tubes}},
  \href{https://doi.org/10.1007/JHEP10(2021)126}{\emph{JHEP} {\bfseries 10}
  (2021) 126} [\href{https://arxiv.org/abs/2106.07957}{{\ttfamily
  2106.07957}}].

\bibitem{Hogervorst:2013sma}
M.~Hogervorst and S.~Rychkov, \emph{{Radial Coordinates for Conformal Blocks}},
  \href{https://doi.org/10.1103/PhysRevD.87.106004}{\emph{Phys. Rev.}
  {\bfseries D87} (2013) 106004}
  [\href{https://arxiv.org/abs/1303.1111}{{\ttfamily 1303.1111}}].

\bibitem{Caron-Huot:2017vep}
S.~Caron-Huot, \emph{{Analyticity in Spin in Conformal Theories}},
  \href{https://doi.org/10.1007/JHEP09(2017)078}{\emph{JHEP} {\bfseries 09}
  (2017) 078} [\href{https://arxiv.org/abs/1703.00278}{{\ttfamily
  1703.00278}}].

\bibitem{Qiao:2017xif}
J.~Qiao and S.~Rychkov, \emph{{A tauberian theorem for the conformal
  bootstrap}}, \href{https://doi.org/10.1007/JHEP12(2017)119}{\emph{JHEP}
  {\bfseries 12} (2017) 119}
  [\href{https://arxiv.org/abs/1709.00008}{{\ttfamily 1709.00008}}].

\bibitem{Mukhametzhanov:2018zja}
B.~Mukhametzhanov and A.~Zhiboedov, \emph{{Analytic Euclidean Bootstrap}},
  \href{https://doi.org/10.1007/JHEP10(2019)270}{\emph{JHEP} {\bfseries 10}
  (2019) 270} [\href{https://arxiv.org/abs/1808.03212}{{\ttfamily
  1808.03212}}].

\bibitem{Bianchi:2021piu}
L.~Bianchi, G.~Bliard, V.~Forini and G.~Peveri, \emph{{Mellin amplitudes for 1d
  CFT}}, \href{https://doi.org/10.1007/JHEP10(2021)095}{\emph{JHEP} {\bfseries
  10} (2021) 095} [\href{https://arxiv.org/abs/2106.00689}{{\ttfamily
  2106.00689}}].

\end{thebibliography}\endgroup
\bibliographystyle{jhep}

\end{document}